\documentclass[aps,shownopacs,superscriptaddress,nofootinbib,preprint,11pt]{revtex4}
\usepackage{hyperref}
\usepackage{amsmath}
\usepackage{float}
\usepackage{graphics}
\usepackage{graphicx}
\usepackage{epsfig}
\usepackage{psfrag}
\usepackage{color}
\usepackage{slashed}

\usepackage{amsfonts}
\usepackage{amssymb}

\newcommand*{\be}{\begin{equation}}
\newcommand*{\ee}{\end{equation}}
\newcommand*{\bea}{\begin{eqnarray}}
\newcommand*{\eea}{\end{eqnarray}}

\newcommand{\comment}[1]{}

\newcommand{\cref}[1]{Chapter~\ref{c.#1}}

\def\beq{\begin{equation}}
\def\eeq{\end{equation}}
\def\bea{\begin{eqnarray}}
\def\eea{\end{eqnarray}}
\def\ba{\begin{array}}
\def\ea{\end{array}}
\def\bi{\begin{itemize}}
\def\ei{\end{itemize}}
\def\be{\begin{enumerate}}
\def\ee{\end{enumerate}}
\def\bc{\begin{center}}
\def\ec{\end{center}}
\def\bt{\begin{table}}
\def\et{\end{table}}
\def\btb{\begin{tabular}}
\def\etb{\end{tabular}}

\def\lsim{\raise0.3ex\hbox{$\;<$\kern-0.75em\raise-1.1ex\hbox{$\sim\;$}}}
\def\gsim{\raise0.3ex\hbox{$\;>$\kern-0.75em\raise-1.1ex\hbox{$\sim\;$}}}

\begin{document}

\title{Lepton Masses and Flavor Violation in Randall Sundrum Model}
\author{Abhishek M Iyer}
\email{abhishek@cts.iisc.ernet.in}
\author{ Sudhir K  Vempati}
\email{vempati@cts.iisc.ernet.in}
\affiliation{Centre for High Energy Physics, Indian Institute of Science,
Bangalore 560012}

\begin{abstract}

Lepton masses and mixing angles via localization of 5D fields  in the bulk are revisited in
the context of  Randall-Sundrum models. The Higgs is assumed to be localized on the IR brane. 
Three cases for neutrino masses are considered: 
(a) The higher dimensional LH.LH operator (b) Dirac masses (c) Type I see-saw
with bulk Majorana mass terms. 
Neutrino masses and mixing as well as charged lepton masses are fit in the first 
two cases using   $\chi^2$ minimisation for the bulk mass parameters, while varying the
$\mathcal{O}(1)$ Yukawa couplings between $0.1$ and $4$. Lepton flavour violation is studied
for all the three cases.  It is shown that
large negative bulk mass parameters are required for the right handed fields to fit the data in the LH LH case. 
This case is characterized by a very  large Kaluza-Klein (KK) spectrum and relatively weak flavour violating constraints 
at leading order.  The zero modes for the charged singlets are  composite in this case
and their corresponding effective 4-D Yukawa couplings to the KK modes  could be large. For the Dirac case, good fits can be obtained for the bulk mass parameters, $c_i$, lying between
$0$ and $1$.  However, most of the `best fit regions' are  ruled out 
from flavour violating constraints.  In the bulk Majorana terms case, we have solved the 
profile equations numerically.  We give example points for inverted hierarchy and normal hierarchy
of neutrino masses.   Lepton flavor violating rates are large for these
points. We then discuss various minimal flavor violation (MFV)  schemes for  Dirac and
bulk Majorana cases.  In the Dirac case with MFV hypothesis, it is possible to 
simultaneously fit leptonic masses and mixing angles and  alleviate lepton flavor 
violating constraints for Kaluza-Klein modes with masses of around 3 TeV.  Similar examples are also provided 
in the Majorana case.

\end{abstract}
\vskip .5 true cm

\maketitle
\section{Introduction}
One of the most interesting solutions of the hierarchy problem is the Randall-Sundrum model \cite{RS} which proposes a \textit{warped}  extra 
space dimension compactified on an $S_1/Z_2$ orbifold. Two branes representing the UV and the IR  scales are located at the two end points of the
orbifold. In the simplest models, the Standard Model matter and  gauge fields are localized on the IR brane along with the Higgs field.  Massive Planck 
scale modes are exponentially suppressed at the IR brane, due to the warped bulk geometry,  caused by the presence of a large negative 
cosmological constant\footnote{ The RS  metric is given by 
 $$ds^2 = e^{-2 \sigma(y) } \eta_{\mu \nu} dx^\mu dx^\nu - dy^2, $$ where $\sigma(y) = k |y|$. 
 For recent reviews on RS models, please see \cite{gherghetta}.}. 
 Variations of this set up have been considered in several different contexts\footnote{The phenomenology of RS models has been extensively studied. A recent review on 
collider  phenomenology concentrating on LHC can be found in \cite{shrihari}.}. 
 
For example, introducing gauge fields in the bulk facilitates unification of couplings~\cite{Agashe:2002pr}. But this leads to large corrections to the 
electroweak  precision observables and places a lower bound on the mass of the lightest gauge Kaluza-Klein (KK) mode to be around 25 TeV. 
This is because the coupling of brane localized fermions to the gauge KK
states is enhanced by a factor $\sim~ 8.5$ compared to the SM coupling~\cite{Davou,Hisano,Huber:2000fh}. A similar study in terms of oblique parameters was
reported in \cite{Csaki:2002gy,Burdman:2002gr}.  Boundary kinetic terms for the gauge fields  can lower the bound \cite{Davoudiasl:2002ua,Carena:2002me},
but this might spoil the unification.  Alternatively, allowing the fermions to propagate in the bulk eases the constraint of 25 TeV on the lightest KK mode,  
to about  10 TeV \cite{Huber:2001gw}.  Having a bulk Higgs further eases the bound \cite{Huber:2000fh}.   On the other hand, scenarios with extended 
particle content and a bulk custodial symmetry with a brane localized Higgs boson were found to lower the bounds on the KK gauge 
boson mass to $\sim \text{3 TeV}$ \cite{Agashe:2003zs}.  In \cite{Hewett:2002fe} the authors explored a mixed scenario where part of fermions, the third
generation quarks  are localized on the IR brane. It was shown that such a scenario would  soften the corrections to the $\rho$ parameter. Finally 
modifying the RS metric near the IR boundary can also help in reduction of the strong electroweak precision constraints \cite{Falkowski:2008fz,Cabrer}.

Allowing fermions to propagate in the bulk has interesting implications for flavor physics.  The bulk profiles of the fermion fields
are  determined by their bulk masses  in a manner similar to   Arkani-hamed and Schmaltz mechanism  in ADD 
models \cite{ArkaniSch}. In the RS model,  however, the warped geometry facilitates the so-called  `automatic' localization of fermions\cite{Hisano}.  
The profiles are also no longer gaussian,  but are exponentially suppressed.  It has been proposed that RS could be a theory of flavour, where the fermion 
 mass hierarchy can be explained in terms of a few $\mathcal{O}$(1) parameters. This is analogous to the popular Froggatt-Nielsen (FN) models \cite{FN,Babu}
 in four dimensions.  While in the FN model, it is the gauge and the heavy fermion sector which determine the hierarchies in the Yukawa
couplings, in the RS case, it is the geometry of the bulk. The role of the FN charges can be played by the five dimensional Dirac masses
for the bulk fermions.  The expectation is that by taking $\mathcal{O}(1)$ bulk  mass parameters as well as Yukawa couplings, one
would be able to explain the large hierarchies in the quark and leptonic mass spectrum.  While this is true in general for quarks and charged 
lepton masses, as we will see subsequently, in case of neutrino masses, the situation is a bit more involved.

Flavour violation in the hadronic sector has been explored by various  authors \cite{AgasheSoni, Huber1,cedric}, a recent comprehensive 
analysis can be found in \cite{neubert1,neubert2}.   In the present work, we are interested in studying neutrino masses and mixing angles within the 
RS context. One method of generating neutrino masses in the RS model would be to allow \textit{only} the right handed neutrino to propagate in the bulk, while
  the SM particles are confined to the IR brane. This leads to a  higher dimensional seesaw mechanism \cite{gross}. However, unlike the case of 
  ADD models, here only the lightest KK modes participate in the  seesaw mechanism. Furthermore, lepton flavour violating decay rates are extremely 
  large in this case pushing the lightest KK mode to be  heavier than $m_{\text{kk}} \gtrsim 25~\text{TeV}$ \cite{Kitano}.  
 Neutrino mass models have also been explored in the 
  alternative scheme  where all the fermionic fields are allowed 
 to propagate in the bulk. 
In the present work, we will concentrate on this set up and  study the  neutrino  mass phenomenology and lepton 
flavor violation  \cite{gross,Kitano,Huber4,Huber3,Huber2,Agashe,Fitzpatrick,Chen,AgasheSundrum}.  
We have assumed  Higgs to be localized on the IR brane. Fermion mass fits in scenarios with Higgs also propagating in the bulk have been considered in
\cite{Huber1,Archer}.
  
 In this RS set-up (fermions in the bulk, Higgs localized on IR brane )  neutrino mass models can be divided broadly 
 into  Dirac mass models or Majorana mass models. In the case of Majorana fermions, 
 the number of possibilities is more than one.  In the present work  we discuss three cases in detail 
 (a) The higher dimensional LH LH operator (b) the Dirac neutrino case and finally (c) Majorana neutrinos with bulk seesaw terms. 
  In  these models,   typically two sets of parameters determine the charged lepton masses and neutrino
masses and mixing angles.  These are the  afore mentioned  set of bulk Dirac masses for the fermions and then the 
$\mathcal{O}(1)$ parameters  containing  the Yukawa couplings.  In each of these cases, we have numerically minimized
a $\chi^2$ function containing the model parameters and the leptonic  masses and mixing data, to determine  the `best fit' 
regions of the parameter space.  The Yukawa couplings are varied from $0.1$ to $4$ whereas the ranges for the 
bulk parameters are judiciously chosen to be as wide as possible. 
 
We found that in the (a) higher dimensional LHLH operator case,  the bulk mass parameters of the charged
singlets  are required to be negative and extremely large. This gets  reflected into an extremely hierarchal Kaluza-Klein mass 
spectrum of the first KK states of the SM fermions. 
In fact, the best fit regions are those with Standard Model charged singlets being completely composite\footnote{This interpretation is based on the AdS/CFT
correspondence.}. On the other hand, if one considers Dirac neutrinos, it is quite possible to fit the data naturally with the bulk Dirac masses 
within reasonable ranges without any large hierarchies. Both hierarchal and inverse hierarchal neutrino mass schemes can be fit in this case
though it is much more difficult to find regions which satisfy inverse hierarchal neutrino mass relations compared to normal hierarchy. 
The bulk equations of motion in the presence of a Majorana mass term  are coupled and more complicated than the Dirac or LHLH case. 
We  have solved them numerically and given example points where data can be fit easily either the inverted or the normal hierarchy scheme. 
 We have not conducted an extended numerical scan of the parameter space for the bulk Majorana case. 

 Fitting neutrino masses in any of the above models in RS set up  potentially leads to large lepton flavor violation.
A detailed analysis  was  presented  in \cite{Agashe}, where the authors discussed the implications
 of  flavor physics in the lepton sector with both the brane localized  and the bulk Higgs. Neutrinos were assumed to be of Dirac nature.
  They observed that with a bulk Higgs, the branching
 fraction for the process $\mu\rightarrow e\gamma$ requires a KK  mass scale of  around $\sim 20$ TeV  to keep it  below the present experimental limits. Similar comments 
  were made in  \cite{AgasheSundrum}  on how the higher dimensional operator case is not conductive for  suppressing process
 like $\mu\rightarrow eee $, especially when the KK mass is low. Higgs was allowed to propagate in the bulk in this work.
  In the present work, we revisited the flavor constraints for
 all the three cases, concentrating on the best fit regions in the LHLH and the Dirac case.  For the LHLH case, the couplings of SM fermions to KK gauge bosons are universal in the best fit region, leading to no apparent constraint, at least at the leading order  from the tree level flavor violating decays. However, there
 are large Yukawa couplings in this model which make it unattractive from perturbation theory point of view. 
 The best fit region of the Dirac case is strongly constrained
 from tree level decays as well as loop induced decays  like $\mu \to e  + \gamma$. 
 In the brane localized Higgs scenario we are considering here, the limits from dipole processes are cut-off dependent. But,  for cut-off values
 close to the first KK mass scale,  the limits are comparably much stronger. 
 For the bulk Majorana case too, the points we have considered display strong constraints from leptonic flavor violation and are 
 ruled out.  One  would thus need ways to circumvent these strong limits from lepton flavor violation.  
 
 We explored Minimal Flavour Violation (MFV) ansatz  implemented in the RS scenario to evade the  flavour constraints
 in the Dirac and Majorana cases~\cite{Fitzpatrick,perez}.  We provide example symmetry groups where the flavor violating constraints can be
 removed for both the Dirac and the Majorana cases. 
       
The paper is organized as follows. In  section (\ref{fitsection}), we discuss lepton mass fits in three models of neutrino
mass generation, the higher dimensional LHLH operator, the Dirac case and the bulk Majorana mass terms
case spread over three subsections. In section (\ref{seclfv}), we discuss the lepton flavor violating constraints for the
three cases of neutrino masses. In section (\ref{secmfv}) we discuss the minimal flavor violating schemes for the
Dirac and Majorana cases and show example points where flavor violating constraints are alleviated. We close
with a summary and outlook in the final section \ref{secfinal}.

\section{Lepton Mass Fits}
\label{fitsection}

The observed neutrino and charged lepton data  is fit to the set of theory parameters which determine the charged lepton and neutrino mass matrices through a  $\chi^2$ minimization. 
 Thus the observables correspond to three charged lepton masses, three mixing angles
 and two (neutrino) mass squared differences, while,  the bulk mass parameters and Yukawa couplings form the set of theory parameters. The number of theory
 parameters varies from model to model,  as discussed in the following sub-sections.  We have chosen the following central values for the observables \cite{pdg,valle}:
 \begin{table}[h]
\caption{Experimental Data}
\begin{center}
\begin{tabular}{|c|c|c|}
\hline 
masses &mass-squared& mixing angles\\
(MeV )& ($\text{eV}^2$)&  \\
\hline
\hline
$m_e = 0.51^{+0.0000007}_{-0.0000007}$&$\Delta m^2_{12} = 7.59  ^{+0.20}_{-0.21} \times 10^{-5} $&$ \theta_{12} =0.59^{+0.02}_{-0.015}$ \\
$m_{\mu} = 105.6^{+0.000003}_{-0.000003}$&$\Delta m_{23}^2 = 2.43^{+0.13}_{-0.13}  \times 10^{-3}    $&$\theta_{23} = 0.79^{+0.12}_{-0.12}$\\
$m_{\tau} = 1776^{+0.00016}_{-0.00016}$&&$\theta_{13}=0.154^{+0.016}_{-0.016}$\\
\hline
\end{tabular}
\end{center}
\label{inputtable}
\end{table}%

We use the standard $\chi^2$ definition for N observables given by 
\begin{equation}
 \chi^2=\sum_{i=1}^N\left(\frac{y_i^{exp}-y_i^{theory}}{\sigma_i}\right)^2
\end{equation}
where, $y_i^{theory}$ is the value of the $i^{th}$ observable predicted by the model and $y_i^{exp}$ is its corresponding experimental number measured with a uncertainty of $\sigma_i$.  Since, the values of the charged lepton are measured to a very high accuracy, it is difficult to fit masses to such high accuracy.
Thus, we incorporate up to $\sim 1.5\%$ errors in the masses of charged leptons\footnote{This approach is very similar to fermion mass fitting in GUT theories. See for example, \cite{GUT1,GUT2}.}.
  The $\chi^2$ relevant to our study is

\begin{eqnarray}
 \chi^2 = \frac{(\theta_{sol}-0.59)^2}{(0.02)^2}+
    \frac{(\theta_{atm}-0.79)^2}{(0.12)^2}+
    \frac{(\theta_{13}-0.154)^2}{(0.02)^2}+
    \frac{(\Delta m^2_{sol}-7.59\times 10^{-23})^2}
    {(0.2\times10^{-23})^2}   \nonumber \\
 +     \frac{(\Delta m^2_{atm}-2.43\times10^{-21})^2}
     {(0.2\times10^{-21})^2}+
    \frac{(m_e-0.00051)^2}{(0.00001)^2} +
     \frac{(m_\mu - 0.1056)^2}{(0.0001)^2} +
    \frac{(m_\tau- 1.77)^2}{(0.02)^2} 
    \label{chisq}
 \end{eqnarray}
 
As mentioned above, the fermion masses (and mass squared differences) and mixing angles appearing in Eq.(\ref{chisq}) are functions of bulk parameters. 
   The minimization was performed using MINUIT~\cite{minuit}.  For a given scan, MINUIT looks for a local minima for the $\chi^2$ around a certain input guess value of the bulk masses and Yukawa parameters. This scan is repeated by randomly varying the 
guess values and in the process of looking for a global minima.  

\subsection{The $LH LH$ operator}
In the absence of  detailed specification of the mechanism which generates neutrino masses, one can always write an 
effective higher dimensional operator at the weak scale to account for non-zero neutrino masses. 
In the Standard Model, this operator is simply the $(LH LH)/ \Lambda$ operator , where $\Lambda$ is the high scale
at which neutrino masses are generated. In the Randall Sundrum model
 one can write a similar operator
   for non-zero neutrino masses.  The model has been earlier studied in \cite{Huber1,Huber3}. The 5D action for the RS model with the Higgs localized on the IR brane is given by 
\begin{eqnarray}
\label{lhlhaction}
S &=& S_{\text{kin}} + S_{\text{Yuk}} \nonumber \\ 
 S_{kin} &=& \int d^4x\int dy ~\sqrt{-g}~ \left(~ \bar{L} (i\slashed D - m_L)L + \bar{E} (i\slashed D - m_{E})E ~\right) \nonumber \\ 
S_{\text{Yuk}}  & =& \int d^4x\int dy~\sqrt{-g} \left(\frac{\mathbf{\kappa}}{\Lambda^{(5)}}LHLH ~+ ~  Y_E \bar{L}EH \right) \delta(y-\pi R) 
\end{eqnarray}
where $\Lambda^{(5)} \sim 2.2 \times 10^{18}$ GeV is the fundamental five dimensional reduced Planck scale and 
\begin{equation}
\label{covariant}
 D_M =  \partial_M + \Omega_M +\frac{ig_5}{2}\tau^aW_{M}^a(x,y) + \frac{ig'}{2}Q_YB_M(x,y) 
\end{equation}
with $\Omega_M =  ( -k/2 e^{- k y} \gamma_\mu \gamma^5, 0)$ being the spin connection and $Q_Y$ is the hypercharge. $M$ is the five dimensional Lorentz index. 
$R$ is the compactification radius and  $\kappa$ and  $Y_E$  are the coupling of the neutrino mass operator and  the Yukawa coupling for the charged leptons respectively. They
  are three dimensional matrices in flavour space and we have suppressed the generation indices in writing the above equation. 
   $L$ and $E$ are the 5D fermionic fields which transform as doublets and singlets respectively under the Standard Model $\text{SU(2)}_\text{W}$ gauge group
   with the covariant derivative given by Eq.(\ref{covariant}) acting accordingly.  $m_L$ and $m_{E}$ are five dimensional  Dirac masses of the $L$ and $E$ fields. As we will
   see below, after Kaluza-Klein decomposition, these masses  
determine the profiles of the zero and higher KK modes in the extra dimension.  Since the effective operator is suppressed by the 5D Planck mass, one can imagine that 
the neutrino masses are as a  result of some fundamental lepton number violation beyond the 5D Planck scale. 

\begin{figure}[htp]
\begin{tabular}{cc}
\includegraphics[width=0.50\textwidth,angle=0]{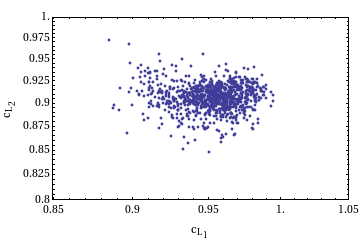} &
\includegraphics[width=0.50\textwidth,angle=0]{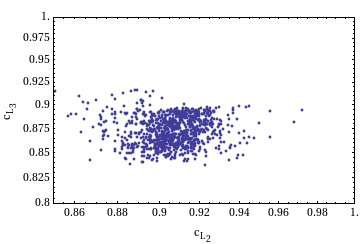} \\
\includegraphics[width=0.50\textwidth,angle=0]{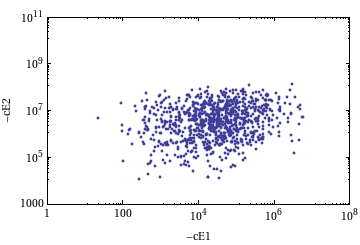} &
\includegraphics[width=0.50\textwidth,angle=0]{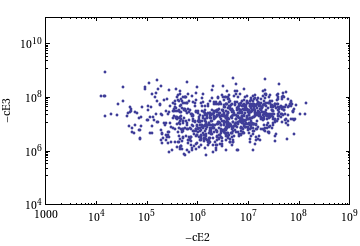}
\end{tabular}
 \caption{Regions in $c_i$ for the LHLH case which give best fit to lepton masses and mixing. The graphs in the upper row shows the region of parameter space for
 the bulk masses for doublets which fits small neutrino masses. Neutrino masses are assumed to have normal hierarchy
 in this analysis. The graphs in the lower row shows the region for the bulk masses for the charged singlets  $c_{E_i}$. We have used log scale for $c_{E_i}$  }
\label{lhlhfit1}
\end{figure}

The left and right components of the $L$ and $E$ fields have different $Z_2$ properties. These are chosen such that the $Z_2$ even
 zero  modes correspond to the SM fields. 
We assign the following $Z_2$ parity for the  $L_{l,r}$ and the $E_{l,r}$ fields, where  the subscript $(l,r)$ 
correspond to the left and right handed components of $L$ and $E$ \footnote{The $\gamma_5$ required
to define the left and right components remains the same as the four dimensional case.}.
\begin{eqnarray}
\label{z2properties}
 Z_2(y) L_l(x,y)\rightarrow L_l(x,y) &,& Z_2(y) L_r(x,y)\rightarrow - L_r(x,y) \nonumber \\
Z_2(y) E_r(x,y)\rightarrow E_r(x,y) &,&Z_2(y) E_l(x,y)\rightarrow - E_l(x,y) \nonumber,
\end{eqnarray}
where $Z_2(y) : y \rightarrow -y$. 
The 5D fields can be expanded in terms of the KK modes, with the expansion given by \cite{gross,AgasheSoni}  
\begin{eqnarray}
\label{kkexpansion}
L_{l} (x, y) = \sum_{n=0}^\infty {1 \over \sqrt{\pi R} } e^{2 \sigma(y) } L_{l}^{(n)}(x) f^{(n)}_{L }(y) &;&L_{r}(x, y)= \sum_{n=0}^\infty {1 \over \sqrt{\pi R} } e^{2 \sigma(y) } L_{r}^{(n)}(x) \chi^{(n)}_{L }(y) \nonumber \\
E_{r}(x,y)= \sum_{n=0}^\infty {1 \over \sqrt{\pi R} } e^{2 \sigma(y) } E_{r}^{(n)}(x) f^{(n)}_{E }(y)&;& E_{l}(x,y)= \sum_{n=0}^\infty {1 \over \sqrt{\pi R} } e^{2 \sigma(y) } E_{l}^{(n)}(x) \chi^{(n)}_{E}(y)
\end{eqnarray} 
where the exponential factor is chosen such that the fields are canonically normalized. The profiles $f_{L,E}$ and $\chi_{L,E}$ are determined by : 
\begin{eqnarray}
\label{eigenvalueeqn}
 (\partial_y + c_{L} \sigma')f_{L,E}^{(n)}(y)=m^{(n)} e^{\sigma (y)} \chi^{(n)}_{L,E}(y) \nonumber \\
 (-\partial_y + c_{L} \sigma')\chi_{L,E}^{(n)}(y)=m^{(n)} e^ {\sigma(y)}  f_{L,E}^{(n)}(y)
\end{eqnarray}
where the 5D masses $m_{L,E}$ are written in terms of the fundamental scale as $m_{L,E}=c_{L,E} \sigma'$ and $\sigma'=\partial_y\sigma = k $. The following orthonormality 
conditions are used for the profiles $f_{L,E}$ and $\chi_{L,E}$ to arrive at Eq.(\ref{eigenvalueeqn})
\begin{equation}
{ 1 \over \sqrt{2\pi R}}\int_{-\pi R}^{\pi R}~dy~ e^\sigma\chi^{(n)}_{L,E }(y)\chi^{(m)}_{L,E }(y)={1 \over \sqrt{2\pi R}}\int_{-\pi R}^{\pi R}~ dy~ e^\sigma f^{(n)}_{L,E }(y)f^{(m)}_{L,E }(y)=\delta^{nm}
\end{equation}

The above equations decouple for the zero mode solutions where $m^{(n)} = 0.$  The solution for the $Z_2$ even part, $f_L(y)$ 
 is given as 
 \begin{equation}
 f_L^{(0)} (y) = N_0(c_L) e^{-c_L\sigma' y} \;\;\;\;;\;\;\;N_0 (c_L) = \sqrt{\pi R}\sqrt{\frac{(1-2c_L)k}{e^{(1-2c_L)k\pi R}-1}}
 \label{profile}
 \end{equation}
 $N_0$ being  the normalization constant.  The solution is the same for profile of $E$, $f_E(y)$, with $c_L$ replaced by $c_E$. 
  The bulk wave functions are exponentials  which peak towards the UV (IR) for $c > 1/2$ ($c<1/2$) as can be seen from Eq.(\ref{profile}).  
 Typically, particles lighter in mass  like leptons require $c>1/2$ whereas heavier particles like top quark is localized
 much closer to the IR brane with $c<1/2$.   For the charged leptons and the neutrino masses one would expect all the corresponding 
 $c_i$ to be $> 1/2$. The KK expansions (\ref{kkexpansion}) are put into the Yukawa part of  the action Eq.(\ref{lhlhaction}) leading to 
\begin{eqnarray}
 S_{\text{Yuk}} &=& \int d^4x\int_0^{\pi R} dy \frac{1}{\pi R} \sum_{n,m} ~\left(~ Y_E\bar L^{(n)}(x)f_{L}^{(n)}(y)E^{(m)}(x)f_{E}^{(m)}(y) e^{kR\pi}  H \right.  \nonumber \\
 &+& \left. \frac{\kappa}{\Lambda^{(5)}} f_L^{(n)}(y) f_L^{(m) } (y) L^{(n)}L^{(m)}HH e^{2kR \pi} ~\right)  \delta(y-\pi R), 
\end{eqnarray}
where we have used $H\rightarrow e^{kR\pi}H$ to canonically normalize the Higgs field and suppressed the subscripts $(l,r)$ for the $Z_2$ even 
fields. The odd fields are neglected as they are removed from the boundary as a consequence of the $Z_2$ symmetry.  The charged lepton mass matrix
and the neutrino mass matrix  are determined when the zero modes of the fields are taken. The charged lepton mass matrix, corresponding to the 
$L^{(0)} E^{(0)} H$ operator in the action  is given by 
\begin{eqnarray}
\label{chargedleptonmass}
{\mathcal M}^{(0,0)}_e &=&\frac{v}{\sqrt{2}}\tilde {Y}_E + \mathcal{O}\Big(f_L^{(0)}(\pi R)\frac{v^3}{M_{KK}^2}f_E^{(0)}(\pi R)\Big) \nonumber \\ \tilde{Y}_E& =& {Y_E \over  R\pi} ~ N_0(c_L) N_0(c_E) ~e^{(1-c_L-c_E)k R \pi},
\end{eqnarray}

\begin{figure}[htp]
\begin{tabular}{ccc}
\includegraphics[width=0.30\textwidth,angle=0]{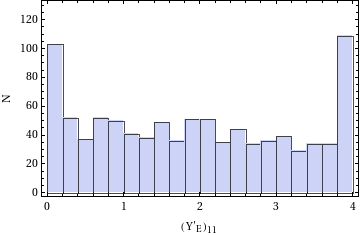} &
\includegraphics[width=0.30\textwidth,angle=0]{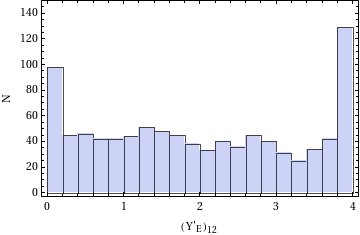} &
\includegraphics[width=0.30\textwidth,angle=0]{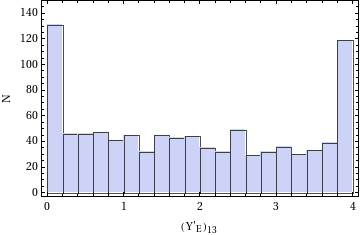} \\
\includegraphics[width=0.30\textwidth,angle=0]{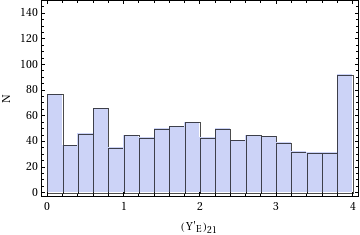}&
\includegraphics[width=0.30\textwidth,angle=0]{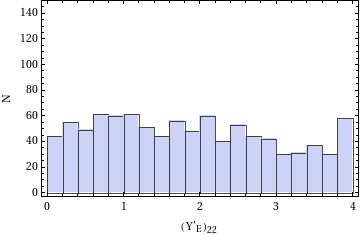}&
\includegraphics[width=0.30\textwidth,angle=0]{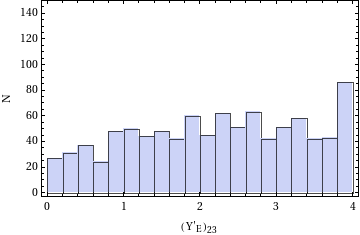} \\
\includegraphics[width=0.30\textwidth,angle=0]{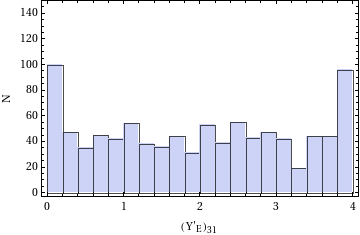}&
\includegraphics[width=0.30\textwidth,angle=0]{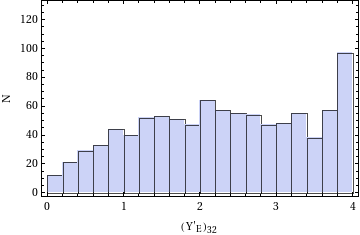}&
\includegraphics[width=0.30\textwidth,angle=0]{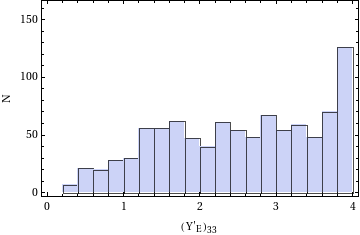} 
\end{tabular}
 \caption{The distribution of electron  Yukawa couplings ($Y_{E}'$) which give a `good fit' to the charged fermion mass data in the LH LH operator case. Neutrinos are assumed to 
follow normal hierarchy in this analysis. The binning is done with an interval of 0.2 }
\label{lhlhyuk1}
\end{figure}
where the matrix $\tilde Y_E$ can be considered equivalent to the 4D dimensionless Yukawa couplings. 
 The neutrino mass matrix defined as
the co-efficient of the $L^{(0)} L^{(0)} H H $ operator in the action, is given as 
\begin{eqnarray}
\label{neutrinomasslhlh}
{\mathcal M}^{(0,0)}_{\nu_{ij}} &=& \tilde\kappa_{ij} \frac{v^2 }{2\Lambda^{(5)}} + \mathcal{O}\left( \frac{1}{M_{KK}} \left(\frac{f_L^{(0)}(\pi R)v^2}{\Lambda^{(5)}} \right)^2 \right) \nonumber \\ 
\tilde\kappa_{ij} &=& \kappa_{ij}  ~ e^{2kR \pi} f_{L_i}(\pi R) f_{L_j} (\pi R) = { \kappa_{ij} \over  R \pi} ~N_0(c_{L_i}) N_0(c_{L_j}) ~ e^{(2-   c_{L_i}-c_{L_j}) kR\pi },
\end{eqnarray}
where $i,j$ are generation indices and $M_{KK}$ is the typical mass of higher KK fermions. The corrections are from higher order KK modes and can be neglected.  
Before fitting the mass matrices, we  introduce new  $\mathcal{O}(1)$ Yukawa parameters entering the
 mass matrices, which  are defined as 
 \begin{equation}
 \label{dimlessyukawa}
 Y_{E}' = 2 k Y_E \;\;\;;\;\; \kappa' = 2 k \kappa
 \end{equation}
In terms of these new Yukawa parameters, the  mass matrices are explicitly given as 
\begin{eqnarray}
\label{massmatrices2}
 ({\mathcal M}^{(0,0)}_e)_{ij} &=&\frac{v}{\sqrt{2}} ({Y}_E')_{ij}  e^{(1-c_L-c_E)k R \pi}    \sqrt{\frac{(0.5-c_{L_i}) }{e^{(1-2c_{L_i})\pi k R}-1}}\sqrt{\frac{(0.5-c_{E_j}) }{e^{(1-2c_{E_j})\pi k R}-1}},
 \nonumber \\ 
 ({\mathcal M}^{(0,0)}_\nu)_{ij} &=&\frac{v^2 }{2\Lambda^{(5)}} (\kappa')_{ij} e^{(2-   c_{L_i}-c_{L_j}) kR\pi }  \sqrt{\frac{(0.5-c_{L_i})}{e^{(1-2c_{L_i})\pi k R}-1}}\sqrt{\frac{(0.5-c_{L_j})}{e^{(1-2c_{L_j})\pi k R}-1}}
\end{eqnarray}

The matrices are diagonalised as $U_{eL}^\dagger \mathcal{M}^{(0,0)}_e U_{eR} = \text{Diag}[\{ m_e,m_\mu,m_\tau\}]$ and $U_\nu \mathcal{M}^{(0,0)}_\nu U_\nu^T = \text{Diag}[\{ m_{\nu_1},m_{\nu_2},m_{\nu_3}\}]$ and $U_{PMNS} = U_{\nu}^\dagger U_{eL} $.  The eigenvalues of the charged lepton mass matrix and the mass squared differences 
of the neutrino mass matrix and the $U_{PMNS}$ mixing angles are fit to the data as per Table \ref{inputtable}. 
In this case, there are three $c_{L_i}$ and three $c_{E_i}$ and fifteen  Yukawa parameters fitting three charged lepton masses, three angles
 and two mass squared differences.  Given the dependence of the leptonic mass matrices on the Yukawa parameters, we have chosen 
 them strictly to be of $\mathcal{O}(1)$ nature. By this we mean, they  are  varied roughly between -4 and 4.  
  Furthermore, in order to avoid regions where the Yukawa parameters 
 are unnaturally close to zero, we put a lower bound on the Yukawas such that $|Y|$ lies between $\sim$ 0.08 and 4. 

\begin{figure}[htp]
\begin{tabular}{ccc}
\includegraphics[width=0.30\textwidth,angle=0]{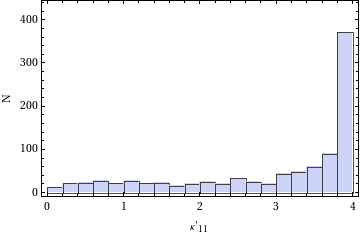} &
\includegraphics[width=0.30\textwidth,angle=0]{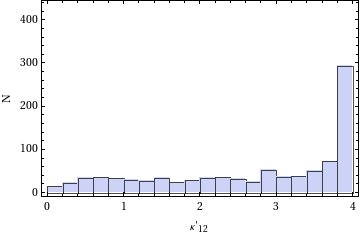} &
\includegraphics[width=0.30\textwidth,angle=0]{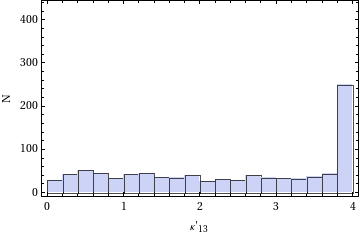} \\
\includegraphics[width=0.30\textwidth,angle=0]{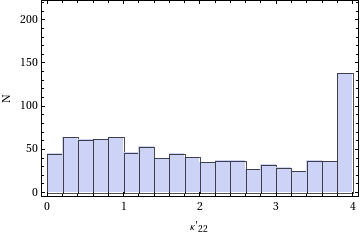}&
\includegraphics[width=0.30\textwidth,angle=0]{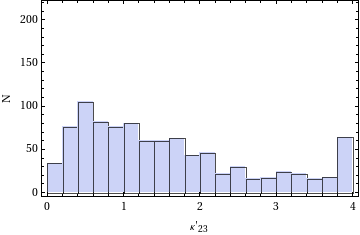}&
\includegraphics[width=0.30\textwidth,angle=0]{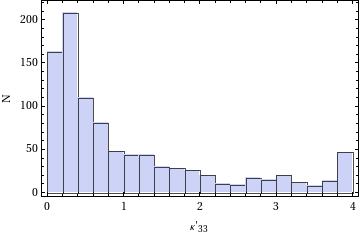} \\
\end{tabular}
 \caption{The distribution of neutrino Yukawa couplings ($\kappa'$) which give a `good fit' to the fermion mass data in the LH LH operator case. Neutrinos are assumed to 
follow normal hierarchy in this analysis. The binning is done with an interval of 0.2.}
\label{lhlhyuk2}
\end{figure}

Since the charged leptons and neutrinos have relatively light mass spectrum compared to heavy quarks, one would have expected
that  varying $c_{L}$ and $c_{E}$ between $1/2$ and $1$ would be sufficient to fit the data. 
However, in the present context such values for
$c_{E}$ will not satisfy the data.   This is because the neutrino mass matrix depends only on $c_{L_i}$ and requiring the neutrino masses
to be of the $\mathcal{O}(10^{-1}) \text{eV}$ automatically sets $c_{L_i}$ to be around $0.9$, close to the UV brane.  
The charged lepton mass matrix,  which in turn is determined by both $c_{L_i}$ and $c_{E_i}$ should off-set the effect of $c_{L_i}$ 
and increase the effective 4D Yukawa coupling by pushing it towards the IR brane. This can only be achieved by taking large and negative 
values\footnote{ One way to avoid large negative c parameters would be to
consider very large O(1) Yukawa parameters. The required Yukawa couplings are in the range $\sim O(10^3-10^4)$ to make any connection with data. 
}
of the $c_{E_i}$. The range for the scan of the $c_{L,E}$  has been judiciously chosen  between
 0.82 and 1.0 for bulk doublets and  $-5 \times 10^{7} <c_{E_1}<-0.2$,
$-10^8<c_{E_2}<-8000$ and $-10^9<c_{E_3}<-9000$ for first, second and third generation charged singlets respectively.
A larger democratic range does not change  the results significantly.

All the parameters, the fifteen Yukawa couplings  and the six $c_{L,E}$ parameters are varied so as to minimize the function in Eq.(\ref{chisq}).  The points 
which give a  $\chi^2$ between 1 and 8 are considered to give a `good fit' to the data.  In Fig.[\ref{lhlhfit1}] we present the regions in
 $c_{L_{1,2,3}}$ and $c_{E_{1,2,3}}$ which have minimum $\chi^2$ assuming normal hierarchy for neutrino masses.  It is important
 to remember that Yukawa couplings are also varied in obtaining this range in the $c_{L,E}$ parameter space.  From the figures
  we see that the strong constraint of neutrino masses limits the $c_{L_i}$ to be within a limited range. On the other hand, 
  $c_{E}$ seem to have much larger ranges spanning orders of magnitudes.  In particular, $c_{E_1}$ 
  is virtually unconstrained from $\mathcal{O}(-1)$ to $\mathcal{O}(-10^{6})$.  This is an artifact of  the unconstrained lightest neutrino 
  mass, $m_{\nu_1}$.  $c_{E_2}$ and $c_{E_3}$ have lesser freedom as they are constrained by the mass squared differences. The allowed ranges 
 in the $c_{L,E}$ which satisfy the minimum $\chi^2$ requirement are summarized  in Table \ref{lhlhnormalrange}.


\begin{table}[htdp]
\caption{Allowed range for the  bulk parameters with minimum $\chi^2$. Neutrino masses have normal hierarchy. Range of first KK scale of the doublet(singlet) $M^{(1)}_{L}$($M^{(1)}_{E}$) corresponding
to the bulk mass parameter is also give.}
\begin{center} 
\begin{tabular}{|c|c|c|c|c|c|}
\hline
parameter & range & range of $M^{(1)}_{L}$ (TeV)&parameter & range& range of $M^{(1)}_{E}$(TeV) \\
\hline
\hline
$c_{L_1}$ & 0.87-0.995 &1.49-1.59 &$c_{E_1}$ & $-10.0$ to $ -5.0\times 10^6$& 7.9-$3.9\times10^6$\\
$c_{L_2}$ & 0.86-0.98 &1.48-1.58 &$c_{E_2} $& $-1.0\times 10^{4}$ to $ -1.2\times 10^{8}$&$7.9\times10^3$-$9.5\times10^7$ \\
$c_{L_3}$ & 0.84-0.92 &1.47-1.53 &$c_{E_3} $& $-7.0\times 10^{5}$ to $-1\times 10^{9}$&$5.5\times10^5$ $7.9\times10^8$ \\
\hline 
\end{tabular}
\end{center}
\label{lhlhnormalrange}
\end{table}

It would be interesting to see distribution of the Yukawa couplings $Y'_E$ and $\kappa'$ for the `best fit' regions of the
parameter space. The distributions are presented in Figs.[\ref{lhlhyuk1}] and [\ref{lhlhyuk2}].  For most of the
$Y'_{E}$ parameters, there is  peaking at the two ends of the range chosen,  around 0.2 and 3.8. The
exception is the lower $2 \times 2$ block of the Yukawa matrix, for which there seems to be a flatter profile for
the upper row parameters $(Y'_{E})_{22} $ and $(Y'_{E})_{23} $ and a progressively increasing distribution
for the second row parameters. 

For almost all 
of the $Y'_{E}$ parameters,  peaking seems to be happening at high values $\sim 3.8$, except for $(Y'_{E})_{22}$. 
 There are also  second peaks at very low values $\sim 0.2$ for 
some of the parameters.  Distributions in $\kappa'$ on other hand, show peak at very large value $\sim 3.8$ 
for the first two generation couplings and very low values $\sim 0.4$ for $\kappa'_{33}$ and $\kappa'_{23}$.  With the exception of peaks,
there is an underlying  though highly subdued, `anarchical' nature in the distribution of $Y'_E$ Yukawa couplings\footnote{Anarchy
in the Yukawa distributions does not necessarily mean anarchical structure in the mass matrix.}. Thus, for a given choice of 
$\mathcal{O}$(1) Yukawa couplings within our chosen range (-4 to 4), it seems to be possible to find $c$ values
which can fit the data well \footnote{Increasing the scan range for the $\mathcal{O}$(1) Yukawa couplings from -10 to 10 does not change
the gross features of the distributions much. For example, $Y'_{E}$ are peaked near the end points, showing that the lepton masses
in this case prefer large or small Yukawa couplings. The $\kappa'$ distribution has the same features scaled now to 
to 0 to 10 from 0 to 4. The ranges of the $c_{L,E}$ do not change significantly.}. 
 From the allowed parameter space, we have randomly chosen two sample points, 
which we call Point A  and Point B,  and we provided the corresponding observables in Table \ref{lhlhobs1}. 
The corresponding Yukawa couplings are given in Eqs. (\ref{yukawa1}) and (\ref{yukawa2}).

\begin{table}[htdp]
\caption{Sample points with corresponding fits of observables for Normal Hierarchy  in LHLH case with $\mathcal{O}(1)$ Yukawas. The masses are in GeV}
\begin{center}
\begin{tabular}{|c|c|c|}
\hline
Point  & A&B\\  
\hline
\hline
$\chi^2$&2.07&5.5 \\
\hline
$c_{L_1}$&0.9755&0.903\\
\hline
$c_{L_2}$&0.9162&0.93\\
\hline
$c_{L_3}$&0.87& 0.8443 \\
\hline
$c_{E_1}$&-692416.99& -17.35\\
\hline
$c_{E_2}$&-2647794.18&-946125.13\\
\hline
$c_{E_3}$&-80717122.21&-47941542.53\\
\hline
$m_e $&$5.07\times10^{-4}$&$5.08\times10^{-4}$ \\
\hline 

$m_\mu$& 0.1056&0.1056\\
\hline
$m_\tau$&1.767&1.771\\                                    
 \hline
    $\theta_{12}$&0.58&0.589\\
\hline
 $\theta_{23}$&0.68&0.743\\
\hline
   $\theta_{13}$&.168&0.163\\
\hline
 $\delta m_{sol}^2$&$7.49\times 10^{-23}$&$7.48\times 10^{-23}$\\
\hline
  $ \delta m_{atm}^2$&$  2.47\times 10^{-21}$&$  1.99\times 10^{-21}$\\                            
\hline 
\hline 
\end{tabular}
\end{center}
\label{lhlhobs1}
\end{table}

\noindent
Yukawa coupling matrices  for Point A:
\begin{equation}
 Y'_E=\begin{bmatrix}
              0.5023       &       1.9546       &       3.9730     \\         3.2482      &        2.9629          &    2.7742       \\       2.6865     &         2.0383      &        1.2369
 \end{bmatrix} \;\;\;; \;\;\; \kappa'=\begin{bmatrix}
              3.8933      &        3.9717       &       3.9818     \\         3.9717      &       -2.6660      &       -1.1409       \\       3.9818        &     -1.1409      &        1.4597
\end{bmatrix}
\label{yukawa1}
\end{equation}

\noindent
Yukawa coupling matrices for Point B
\begin{equation}
 Y'_E=\begin{bmatrix}
              3.0571    &          0.6316      &        0.8978     \\         1.4085         &     0.9952         &     3.5597         \\     0.7971     &         0.9579       &       0.5539
 \end{bmatrix} \;\;\;; \;\;\; \kappa'=\begin{bmatrix}
              0.2315     &        -3.8320      &        0.3490     \\        -3.8320    &         -0.6632       &      -1.1287       \\       0.3490     &        -1.1287         &     0.0802
\end{bmatrix}
\label{yukawa2}
\end{equation}
  In Appendix \ref{invertedfits} we have presented our results assuming
neutrinos have an inverse hierarchical mass ordering. We find very few points which satisfy the data in this case. This is because
inverted hierarchical spectrum requires two masses at the atmospheric neutrino scale with their mass difference satisfying $\Delta m^2_{sol}$. Thus the results are very sensitive
to the $\mathcal{O}$(1) Yukawa parameters. For a fixed Yukawa, however it is easy to find points. More discussion is present in Appendix \ref{invertedfits}.

The analysis presented so far  has been purely phenomenological. Let us digress from the fermion fits for a moment to discuss about the large negative $c$ parameters. 
Such large negative values for the bulk mass parameters are in conflict with the 5D cutoff scale $k$. We have neglected this conflict
in fitting the data where we have considered them to be purely phenomenological parameters which can take any value\footnote{We prefer to keep the $k$ (and also the
radius R) value fixed by noting that
only the charged singlets required large negative $c$ values. In case we shift the 5D cut-off scale to $|c|k$ keeping $k$ fixed, the corresponding
IR would shift to $c\Lambda_{IR}$, thus spoiling the solution to the hierarchy problem in this scenario.}. In terms of the bulk wave-functions
the large negative $c$ values would mean that the zero mode wave-function $f^{(0)} \gg 1$, which is not the case when we choose the
$c$ parameters between 0 and 1.

It is preferable to understand the large negative $c$ values in terms of localization on the IR brane. The limit $c\rightarrow -\infty$ corresponds to the case where the fermions are completely localized
on the IR singular point\cite{Gherghetta:2003he}. In the limit $c\rightarrow -\infty$, $f^{(0)}\rightarrow \infty$ indicating full overlap of the bulk wave-function
with the brane. The value of the $c$ parameters also affects the masses of the KK modes. These masses are determined from Eqs.(\ref{eigenvalueeqn}) by considering
$m_n\neq 0$ and choosing appropriate boundary conditions for the 5D fields. The resultant differential equation has solution
in terms of Bessel's function which describe bulk wave-functions of the KK modes whereas the masses are given in terms of the zeros
of the Bessel function\cite{gher}. The order of the Bessel function is roughly given by $|c|$ for large values of c. In the asymptotic limit the first KK mode has mass 
$\thickapprox |c|ke^{-kR\pi}$. Thus we see that the phenomenologically relevant first KK mode mass also grows as $\sim c\Lambda_{IR}$, where 
$\Lambda_{IR}\sim TeV$, the IR cutoff. The masses of the first KK modes are presented in the Table[\ref{lhlhnormalrange}]. The bulk wave-function
of the KK mode tend to zero as $|c|\rightarrow\infty$.

One might wonder if such large negative values of the $c_{E_i}$ parameters
would have some implications in terms of the AdS/CFT correspondence\cite{randallporrati,gherghetta}.  
The CFT interpretation for the bulk scalars  has been studied in  \cite{batellgherghetta1,gherghetta} and for bulk fermions 
in \cite{continopomarol}.  The best fit $c_{L,E}$ parameters of  LLHH case  given in Table \ref{lhlhnormalrange} leads to
an unusual situation where the left handed leptons are almost completely elementary while the right handed singlets are completely composite. 
This can be easily verified  using the `holographic basis' of \cite{batellgherghetta2}.  The composite component of the $c_{L}$ is proportional to
$e^{-(c_L-0.5)kR\pi}$, which goes to 0 when $c_{L} \to 0.99$. Thus, the zero modes for the doublets are elementary. For the $c_{E}$ fields however, the elementary component for the zero mode is given as $\sqrt{(c_E-0.5)(c_E+1.5)}e^{-|1.5+c_E|kR\pi }$.
 Thus we see that the zero mode for the charged singlets have a vanishing elementary component and are completely composite fields.
The effective 4-D Yukawa coupling of the zero mode to the KK modes, is given as $Y'_E\sqrt{(0.5-c_E)}$. A problematic feature of these models is 
that this coupling enters the non-perturbative
regime for $c_E$ large and negative. This non perturbative coupling appears for all including the first KK mode, which is phenomenologically
relevant. This, non-perturbative feature is restricted to the Yukawa coupling. The gauge coupling on the other hand do not face this
problem. In fact as we shall see later (Section V, Figure[\ref{overlap}]), the coupling strength of the zero mode fermions to gauge KK modes
quickly approaches the coupling of the brane localized fermions to gauge bosons for relatively moderate values of $|c|$ parameter.

\begin{figure}[H]
\begin{tabular}{cc}
\includegraphics[width=0.5\textwidth,angle=0]{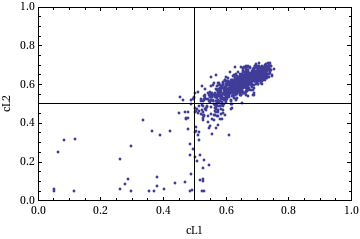} &
\includegraphics[width=0.5\textwidth,angle=0]{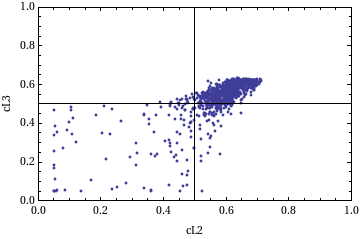} \\
\end{tabular}
 \caption{The figures above correspond to the case in which neutrinos are of Dirac type. The points in the above figures correspond to a $\chi^2$ between 1 and 8. The plot
represents the parameter space for the bulk masses of the  doublets. This case corresponds to the normal hierarchial case. 
}
\label{diracnor1}
\end{figure}

\subsection{Dirac Neutrinos }
Dirac neutrino mass models in the RS setting have been extensively studied in the literature \cite{Agashe}. In \cite{AgasheSundrum}, the authors talked about the 
difficulty of fitting neutrino masses and mixing angles in the same scenario as quarks. Their argument drew inspiration from the fact that 
neutrino mixing angles are anarchic in nature. To address this issue they had a bulk Higgs, with the profile 'sufficiently peaked' near the IR brane and 
introduced a `switching behaviour' to fit the  both charged fermion and the neutrino masses and mixing angles.
We, on other hand, approach this problem in the same way as we have  done in the LHLH case of the previous section. 
We look for regions in the parameter space of the bulk masses which give `good' fits  for a reasonable choice of $\mathcal{O}$(1) Yukawa  couplings. 
The particle spectrum 
of the Standard Model is extended by adding singlet right handed neutrino.  Global lepton number is assumed to be conserved. It
can be violated by quantum gravity effects which manifest at the 5D Planck scale. However, for most of the present analysis, we require lepton number violation
present to be highly suppressed.
 
\begin{figure}[H]
\begin{tabular}{cc}
 \includegraphics[width=0.50\textwidth,angle=0]{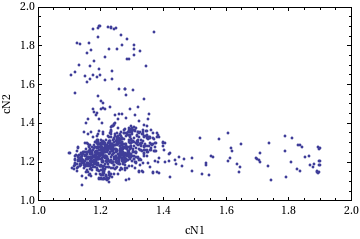}&
 \includegraphics[width=0.50\textwidth,angle=0]{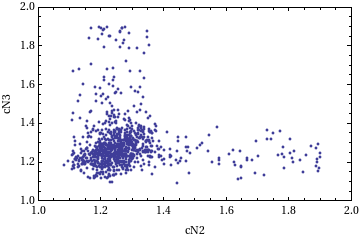}
\end{tabular}
 \caption{The plot represents the parameter space for the bulk masses of the neutrino singlets.}
\label{diracnor3}
\end{figure}

 The bulk and Yukawa actions in Eq.(\ref{lhlhaction}) now take the form: 
\begin{eqnarray}
 S_{kin} &=& \int d^4x\int dy  \sqrt{-g} \left(~~\bar L(i\slashed D - m_L)L + \bar E (i\slashed D - m_{E})E +\bar N (i\slashed D - m_{N}) N ~~\right) \nonumber \\
 S_{yuk}&=&  \int d^4x\int dy~\sqrt{-g} \left(~~  Y_N \bar{L} N H + ~  Y_E \bar{L}EH ~\right) \delta(y-\pi R) ,
  \label{diracaction}
\end{eqnarray}
\begin{figure}[H]
\begin{tabular}{cc}
 \includegraphics[width=0.50\textwidth,angle=0]{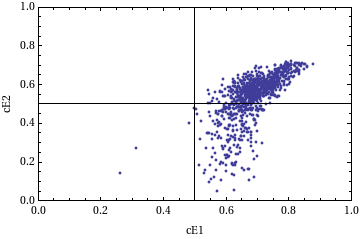}&
 \includegraphics[width=0.50\textwidth,angle=0]{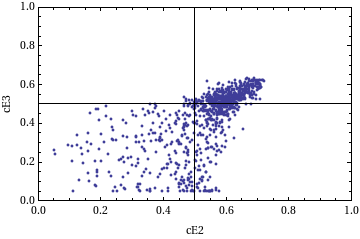}
\end{tabular}
 \caption{The plot represents the parameter space for the bulk masses of charged singlets.}
\label{diracnor2}
\end{figure}
where $N$ stands for the 5D  right handed neutrino fields. The rest of the parameters carry  the same meaning as in the previous
section. The components of the $N$ field are assigned the same $Z_2$ properties as the $E$ field. We expand the $N$ fields as 
\begin{eqnarray}
\label{kkexpansion1}
N_{r}(x,y)= \sum_{n=0}^\infty {1 \over \sqrt{\pi R} } e^{2 \sigma(y) } N_{r}^{(n)}(x) f^{(n)}_{N }(y)&;& N_{l}(x,y)= \sum_{n=0}^\infty {1 \over \sqrt{\pi R} } e^{2 \sigma(y) } N_{l}^{(n)}(x) \chi^{(n)}_{N }(y)
\end{eqnarray} 
Using Eq.(\ref{kkexpansion1}) and Eq.(\ref{kkexpansion}) one can derive the equations of motion  and solutions similar to  Eq.(\ref{profile}) for the 
profiles of $N$ fields.  Substituting them, the zero mode mass matrices for the charged lepton and  neutrinos take the form: 
\begin{eqnarray}
\label{diracleptonmasses}
{\mathcal M}_e^{(0,0)} &=&\frac{v}{\sqrt{2}}\tilde {Y}_E \;;\; \tilde{Y}_E = {Y_E \over  R\pi} ~ N_0(c_L) N_0(c_E) ~e^{(1-c_L-c_E)k R \pi} \nonumber\\
{\mathcal M}_\nu^{(0,0)} &=&\frac{v}{\sqrt{2}}\tilde {Y}_N \;;\; \tilde{Y}_N = {Y_N \over  R\pi} ~ N_0(c_L) N_0(c_N) ~e^{(1-c_L-c_N)k R \pi},
\end{eqnarray}
where we have neglected corrections from higher KK modes. 
As before, we perform a scan over the parameter space of the bulk fermion masses and order one Yukawa parameters to minimize the $\chi^2$ in Eq.(\ref{chisq})  for the 
masses and mixing angles.  To specify the parameters which are scanned, it is useful to look at the explicit form of the mass matrices equivalent to those of Eq.(\ref{massmatrices2}): 
\begin{eqnarray}
\label{mmdirac2}
 ({\mathcal M}^{(0,0)}_e)_{ij} &=&\frac{v}{\sqrt{2}} ({Y}_E')_{ij} e^{(1-c_{L_i}-c_{E_j})k R \pi}  \sqrt{\frac{(0.5-c_{L_i}) }{e^{(1-2c_{L_i})\pi k R}-1}}\sqrt{\frac{(0.5-c_{E_j}) }{e^{(1-2c_{E_j})\pi k R}-1}}\nonumber \\ 
 ({\mathcal M}^{(0,0)}_\nu)_{ij} &=&\frac{v }{\sqrt{2}} ({Y}_{N}')_{ij} e^{(1-c_{L_i}-c_{N_j})k R \pi} \sqrt{\frac{(0.5-c_{L_i})}{e^{(1-2c_{L_i})\pi k R}-1}}\sqrt{\frac{(0.5-c_{N_j})}{e^{(1-2c_{N_j})\pi k R}-1}},
\end{eqnarray}
where  $Y'_{E,N} = 2 k Y_{E,N}$.  Each of the $c_i$ parameters ($ i = \{L,N,E\}$) which are three in number are varied along with eighteen  $\mathcal{O}(1)$ 
 Yukawa parameters, \textit{i.e,} a total of 27 parameters are varied to fit the data and minimize the $\chi^2$.  The $c$ parameters
 are varied as follows:   The doublets ($c_{L_i}$) and the the charged singlets are varied between 0.02 and 1, while the neutral singlets are varied between  between 1 and 1.9.
 The order one Yukawa couplings, $Y'_{E,N}$, are varied randomly between -4 and 4 with a lower bound $|Y| \gtrsim 0.08$. 
We consider all the  regions of the $c_{i}$ parameter space where the $\chi^2$ is between 1 and 8 as a `good' fit region.  
 In Figs.[\ref{diracnor1},\ref{diracnor2},\ref{diracnor3}] we present regions in the $c_i$ parameter space which give `good' fit to the
 leptonic mass and mixing angles. A summary of these regions is presented in Table (\ref{diracnormalrange}). 

\begin{table}[htdp]
\caption{Allowed ranges of bulk parameters with normal hierarchy of neutrino masses. The range of first KK scale corresponding to the range of c values is also given.}
\begin{center}
\begin{tabular}{|c|c|c|c|c|c|c|c|c|}
\hline
parameter & range &$M_L^{(1)}$ TeV & parameter & range& $M_E^{(1)}$ TeV&parameter&range&$M_\nu^{(1)}$ TeV \\
\hline
\hline
$c_{L_1}$ & 0.05-0.76 &0.839-1.4 &$c_{E_1}$ &0.2-0.88&0.959-1.5 &$c_{N_1}$&1.1-1.9&1.67-2.31  \\
$c_{L_2}$ & 0.05-0.72 &0.839-1.37 &$c_{E_2} $&0.05-0.73&0.839-1.38 &$c_{N_2}$ &1.1-1.9&1.67-2.31 \\
$c_{L_3}$ & 0.05-0.64&0.839-1.31 &$c_{E_3} $&0.05-0.64&0.839-1.31 &$c_{N_3}$&1.1-1.9&1.67-2.31   \\
\hline 
\end{tabular}
\end{center}
\label{diracnormalrange}
\end{table}

\begin{figure}[htp]
\begin{tabular}{ccc}
\includegraphics[width=0.30\textwidth,angle=0]{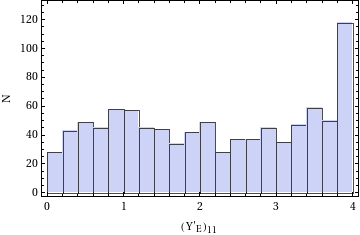} &
\includegraphics[width=0.30\textwidth,angle=0]{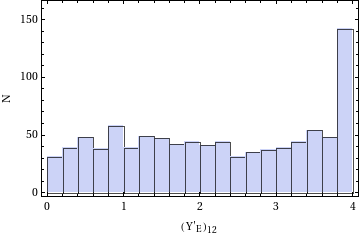} &
\includegraphics[width=0.30\textwidth,angle=0]{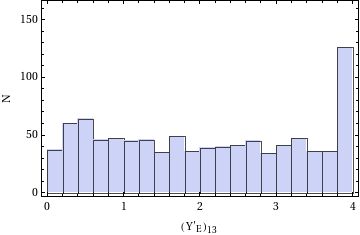} \\
\includegraphics[width=0.30\textwidth,angle=0]{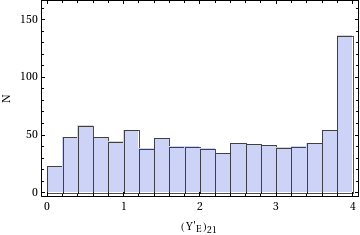}&
\includegraphics[width=0.30\textwidth,angle=0]{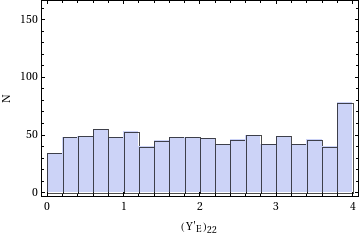}&
\includegraphics[width=0.30\textwidth,angle=0]{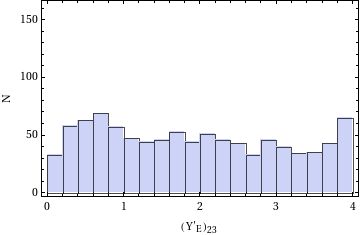} \\
\includegraphics[width=0.30\textwidth,angle=0]{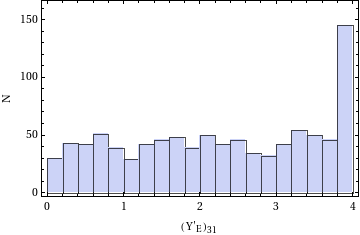}&
\includegraphics[width=0.30\textwidth,angle=0]{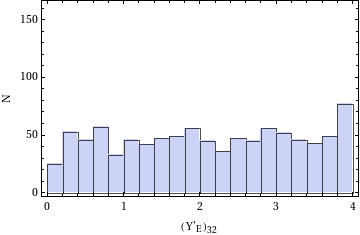}&
\includegraphics[width=0.30\textwidth,angle=0]{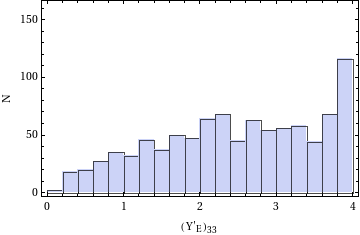} 
\end{tabular}
 \caption{The distribution of electron  Yukawa couplings ($Y_{E}'$) which give a `good fit' to the fermion mass data in the  Dirac case. Neutrinos are assumed to 
follow normal hierarchy in this analysis. The binning is done with an interval of 0.2 }
\label{diracyuk1}
\end{figure}

\begin{figure}[htp]
\begin{tabular}{ccc}
\includegraphics[width=0.30\textwidth,angle=0]{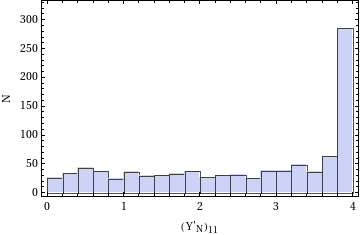} &
\includegraphics[width=0.30\textwidth,angle=0]{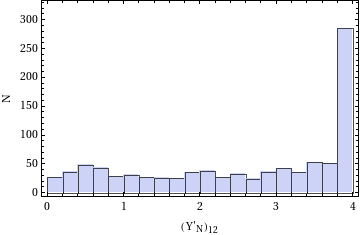} &
\includegraphics[width=0.30\textwidth,angle=0]{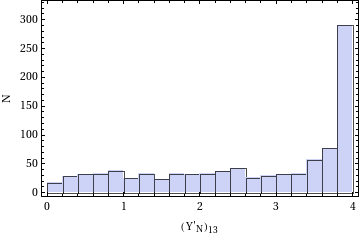} \\
\includegraphics[width=0.30\textwidth,angle=0]{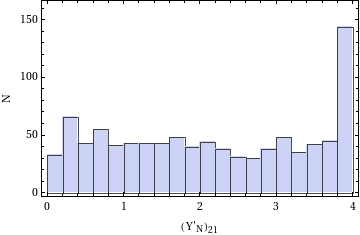}&
\includegraphics[width=0.30\textwidth,angle=0]{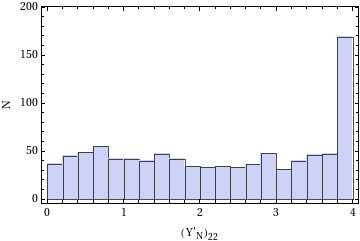}&
\includegraphics[width=0.30\textwidth,angle=0]{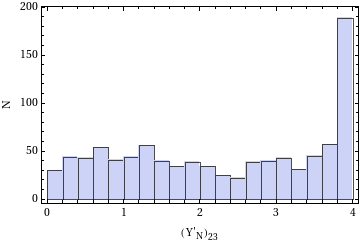} \\
\includegraphics[width=0.30\textwidth,angle=0]{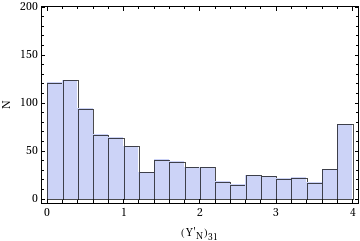}&
\includegraphics[width=0.30\textwidth,angle=0]{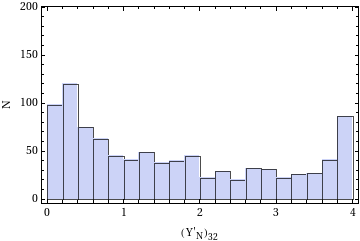}&
\includegraphics[width=0.30\textwidth,angle=0]{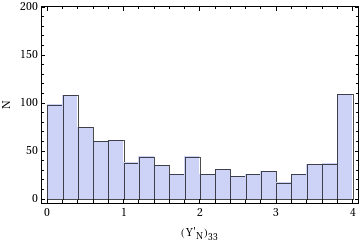} 
\end{tabular}
 \caption{The distribution of neutrino Yukawa couplings ($Y_{N}'$) which give a `good fit' to the fermion mass data in the Dirac case. Neutrinos are assumed to 
follow normal hierarchy in this analysis. The binning is done with an interval of 0.2 }
\label{diracyuk2}
\end{figure}

The Dirac neutrino mass matrix in the RS model seems to fit the data more naturally compared to the $LHLH$ discussed in the previous subsection. 
A large section of the points fall in the regime  $c_i$  $> 1/2$ indicating that they are localized closer to the UV brane.  The distributions of the
Yukawa couplings in the `good fit' region, presented in Figs.(\ref{diracyuk1},\ref{diracyuk2})
show that most of them peak in the  last bins for all the Yukawas at ($3.8-4.0$).  A secondary peak can 
also been seen at  $(0.2-0.4)$ bin for some of the $Y'_{N}$ parameters. Electron Yukawa couplings on the other hand do not seem to show any
such secondary peak. In this case too the distribution of the $\mathcal{O}$(1) Yukawa couplings
displays an underlying  anarchic nature especially for the $Y'_E$. This will prove useful
in our analysis of Minimal Flavour violation where the $\mathcal{O}$(1) Yukawa couplings and the  bulk mass matrices need to
be simultaneously diagonalizable.  In Table(\ref{diracnormal}), 
we presented two sample points. Point A  has all the $c_i > 1/2$ where as Point B has $c_{E_2}, c_{E_3}~<1/2$.  The corresponding Yukawa
couplings are given in Eqs.(\ref{ptadirac},\ref{ptbdirac}). 

As before we use the holographic basis to comment on the partial compositeness of the bulk fermions.  The zero modes of singlet right handed neutrinos are dominantly
elementary, with almost zero component of compositeness. The composite component for the zero modes of the doublets and the charged singlets becomes smaller as the  corresponding c values becomes greater than 0.5. Essentially they have partially composite nature.

\begin{table}[h]
\caption{Sample points with corresponding fits of observables for Normal Hierarchy  in Dirac case with O(1) Yukawas. The masses are in GeV}
\begin{center}
\begin{tabular}{|c|c|c|}
\hline
Parameter &Point A&Point B\\
\hline
\hline
$\chi^2$&0.28&0.39\\
\hline
$c_{L_1}$ & 0.6263&0.7166\\ 
\hline
 $c_{L_2}$ &0.5932& 0.6382 \\
\hline
 $c_{L_3}$&  0.5293&0.6126\\
\hline
$c_{E_1}$& 0.6704&0.5911\\
\hline
$c_{E_2}$ &0.5541&0.1939\\
\hline
$c_{E_3}$&0.5131& 0.2647\\
\hline
$c_{N_1}$&1.2233&1.2791\\
\hline
$c_{N_2}$&1.2692&1.1215\\
\hline
$c_{N_3}$&1.2948&1.2343\\
\hline
$m_e$&$5.09\times10^{-4}$&$5.09\times10^{-4}$\\
\hline
$m_\mu$&0.1055&0.1055\\
\hline
$m_\tau$&1.77&1.77\\
\hline
$\theta_{12}$&0.59&0.589\\
\hline
$\theta_{23}$&0.80&0.792\\
\hline
$\theta_{13}$&0.153&0.153\\
\hline
$\delta m_{sol}^2$&$7.49\times10^{-23}$&$7.49\times 10^{-23}$\\
\hline
$ \delta m_{atm}^2$&$2.39\times 10^{-21}$&$2.40\times 10^{-21}$\\

\hline
\hline
 \end{tabular}
\end{center}
\label{diracnormal}
\end{table}

\noindent 
Yukawa Coupling Matrix for Point A:
\begin{equation}
 Y'_E=\begin{bmatrix}
              3.9502      &       -1.6538       &       0.5889      \\       -0.7276      &       -2.0054       &      -3.9004       \\      -1.4061    &          1.4756      &        1.5318
 \end{bmatrix} \;\;\;; \;\;\; Y_N'=\begin{bmatrix}
             -3.8918     &        -3.9447      &       -3.8380       \\      -2.6439     &         2.5796       &       3.9962        \\     -0.9223     &        -1.3577      &        0.6417

\end{bmatrix}
\label{ptadirac}
\end{equation}

\noindent
Yukawa Coupling Matrix for Point B:
\begin{equation}
 Y'_E=\begin{bmatrix}
              3.3847    &          1.8639     &        -1.3814       \\      -1.8107      &       -0.7219     &        -0.9499        \\     -2.5435        &     -1.0497     &        -3.3588
 \end{bmatrix} \;\;\;; \;\;\; Y_N'=\begin{bmatrix}
              2.4435     &        -1.8006       &      -1.9575    \\          0.4198      &       -3.1594      &        3.5905    \\         -0.2505      &        1.3172     &         2.1521
\end{bmatrix}
\label{ptbdirac}
\end{equation}

 \subsection{Bulk Majorana mass term} 
\begin{figure}[htp]
 \includegraphics[width=0.5\textwidth,angle=0]{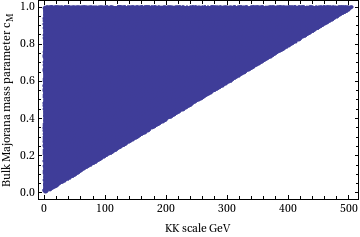}
\caption{Region of $c_M$-$m_1$ parameter space, for positive $c_M$ for which the coefficients of the differential equation in Eq.(\ref{secorder}) are analytic in the interval $[0,\pi R]$}
\label{allowed-maj}
\end{figure}

\begin{figure}[htp]
\begin{tabular}{cc}
\includegraphics[width=0.50\textwidth,angle=0]{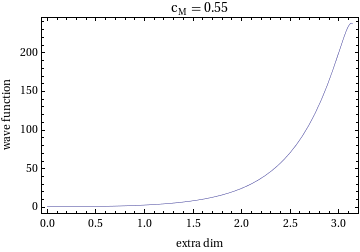} &
\includegraphics[width=0.50\textwidth,angle=0]{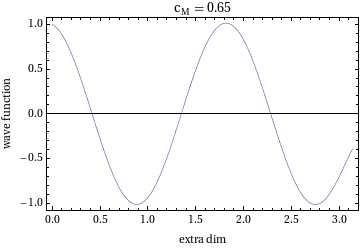}  \\
\includegraphics[width=0.50\textwidth,angle=0]{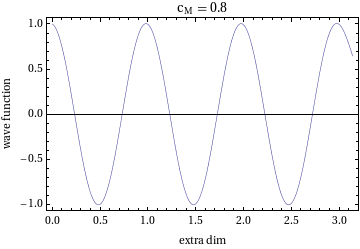} &
\includegraphics[width=0.50\textwidth,angle=0]{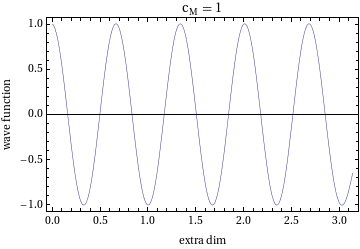}
\end{tabular}
\caption{The Figure shows the form of the profile for solution to Eq.(\ref{secorder}) for a fixed bulk dirac mass of 0.58 for the right handed neutrinos.
We see that the profile becomes oscillatory as $c_M$ becomes greater than $c_N$.}
\label{profile-maj}
\end{figure}

Singlet neutrinos typically  accommodate  Majorana mass terms in addition to the Dirac mass terms. These bare mass terms which break 
lepton number at a very high scale play an essential role in the  standard four dimensional seesaw mechanism to generate light neutrino masses.  
 The seesaw mechanism with bulk Majorana mass terms has been first considered in \cite{Huber2}. There have been 
 other works which have considered brane localised Majorana mass terms \cite{Goldberger:2002pc,Nomura:2003du,Gherghetta:2003he,perez}. Our analysis follows the work of \cite{Huber2} and
 extends it by computing the numerical solutions.  The  part of the action which contains the singlet right handed neutrinos is given by 
\begin{equation}
 S_{N}=\int d^4x\int dy \sqrt{-g} \big(m_M\bar NN^c+ m_D \bar N N + \delta(y-\pi R)Y_N\bar L\tilde H N\big)
\label{majorana}
\end{equation}
where $N^c = C_5 \bar{N}^{T}$ with $C_5$ being the five-dimensional charge conjugation matrix\footnote{$C_5$ is taken to be $C_4$.} and $m_M =c_M k$, with $k$ 
being the reduced Planck scale\footnote{Majorana mass terms
does not have the same interpretation in the bulk as in 4D.}. The bulk Dirac mass for the right handed neutrino is parametrized as $m_D=c_Nk$. As before we consider all the  mass parameters to be real. The bulk singlet fields N have the following KK expansions:
\begin{eqnarray}
\label{kkexpansion2}
N_{L} (x, y) = \sum_{n=0}^\infty {1 \over \sqrt{\pi R} } e^{2 \sigma(y) } N_{L}^{(n)}(x) g^{(n)}_{L }(y) &;&N_{R}(x, y)= \sum_{n=0}^\infty {1 \over \sqrt{\pi R} } e^{2 \sigma(y) } N_{R}^{(n)}(x) g^{(n)}_{R }(y), \nonumber \\
\end{eqnarray}
where $g_L$ and $g_R$ are profiles of the singlet neutrinos in the bulk. They follow the following  orthonormal conditions 
\begin{equation}
 \frac{1}{2\pi R}\int_{-\pi R}^{\pi R}~dy~ e^{\sigma}\Big(g_L^{(n)}g_L^{(m)}+g_R^{(n)}g_R^{(m)}\Big)=\delta^{(n,m)}
\label{bulkmajnormalization}
\end{equation}
Using this, the eigenvalues equations for the  $g_{L,R}$ fields  become \cite{Huber2}
\begin{eqnarray}
(\partial_y+m_D)g_L^{(n)}(y)=m_ne^\sigma g_R^{(n)}(y)-m_Mg_R^{(n)}(y) \nonumber \\
(-\partial_y+m_D)g_R^{(n)}(y)=m_ne^\sigma g_L^{(n)}(y)-m_Mg_L^{(n)}(y)
\label{majcoupled}
\end{eqnarray}
where we have assumed the five dimensional  wave functions to be real. 
Unlike the Dirac and higher dimensional LHLH term cases, the present system  of equations, in Eq.(\ref{majcoupled})  
are not consistent with a zero mode solution $m_n=0$ for $m_D\neq 0$. This
is because the zero mode solutions, $\propto e^{\sqrt{c_N^2-c_M^2}\sigma}$ do not satisfy either Dirichlet or the more general  
$(\partial_y+m_d)g_L(y)=0 $ boundary condition. Thus in the following analysis, we will consider the first KK mode 
not to be the zero mode but $m_n = m_{(1)} $. 
Furthermore,  Eq.(\ref{majcoupled}) does not have simple analytical solutions, though numerical solutions exist.  We 
have obtained the numerical solutions of $g_{L,R}$ by solving the second order equations derived from Eq.(\ref{majcoupled}).
The equation for the $Z_2$ even part takes the form:
 \begin{equation}
 g_L''(y)-\frac{m_nkRe^{kRy}}{m_ne^{kRy} - c_M k }g_L'(y)-\left(\frac{c_Nm_ne^{kRy}k^2}{m_ne^{kRy} - c_M k } + c_N^2k^2-\left(m_ne^{kRy} -c_Mk \right)^2\right) R^2g_L(y) = 0
\label{secorder}
\end{equation}
The second order equation for the $Z_2$ odd part $g_R$ is given as
\begin{equation}
 g_R''(y)-\frac{m_nkRe^{kRy}}{m_ne^{kRy} - c_Mk }g_R'(y)-\left(\frac{-c_Nm_ne^{kRy}k^2}{m_ne^{kRy} - c_Mk } + c_N^2k^2-(m_ne^{kRy} - c_Mk )^2\right)R^2g_R(y) = 0,
\label{secorder1}
\end{equation}
where we have used the notation $m_D = c_Nk$ and $M_M = c_Mk $ introduced earlier. The primes on $g_L(y)$ and $g_R(y)$ indicate
derivatives on the profiles. For a given choice of $c_N$ and $c_M$ one
would expect to numerically find solutions using the above equations for $g_{L,R}$ as long as they satisfy two conditions:
(i) $m_{(1)}$ is also fixed such that the boundary conditions are satisfied consistently (ii) There are no singularities in coefficients
of the differential equations in the interval $[0,\pi R]$. This second condition requires that for unique solutions, only those values
of $c_M$ and $m_n$ are allowed for which $m_ne^{\sigma}-m_M$ is non zero. Note that this condition is always true when $c_M$ is
negative. For positive $c_M$, the allowed region is shown in  Fig.[\ref{allowed-maj}], where all the shaded region has 
$m_ne^{\sigma}-m_M$  non zero. As can be seen from the figure, as $c_M$ increases, the KK mass scale also increases.
In Fig.[\ref{profile-maj}] we show  solutions to Eq.(\ref{secorder}) for a fixed value of  $c_N = 0.58$. $c_M$ is varied from 0.55 to 1.  From the figure,
it is clear as the  profile becomes oscillatory as $c_M$ becomes greater than $c_N$.  In fact the  solutions are sinusoidal for $c_M$=1 and $c_N$=0.
We now address the question of fitting the lepton masses and mixing. 
The charged lepton mass matrix has the same form  as in earlier sections
\begin{equation}
 m_l^{(0,0)} =\frac{v}{\sqrt{2}}\tilde {Y}_E + O(\frac{v^2}{M_{KK}^2}) \;;\; \tilde{Y}_E = {Y_E \over  R\pi} ~ N_0(c_L) N_0(c_E) ~e^{(1-c_L-c_E)k R \pi}.
\end{equation}
Choosing $g_L^{(1)}$ to be the $Z_2$ even profile for the right handed neutrino, the Dirac mass matrix takes the form
\begin{equation}
 m^{(0,1)}_D={Y_N \over  R\pi}N_0(c_L)e^{(1-c_L)kR\pi}g_L^{(1)}(\pi R)
\end{equation}
where $g_L^{(1)}(y)$ is the solution to Eq.(\ref{secorder}). The singlet Majorana mass matrix is determined in the flavor space by the
choice of $c_{N}$ and $c_M$ for each of the generations.  For simplicity, for the present analysis, we take all of them equal $c_{N_i} = c_{N}$
and $c_{M_i} = c_M$ for all the three generations\footnote{This can be achieved by imposing an $O(3)$ symmetry on the $N$ fields.}. With this, singlet neutrino mass matrix becomes proportional to the unit matrix $M_R=\textbf{1}~m_{(1)}$.
 The light neutrino mass matrix now takes the see-saw form given by
\begin{equation}
 m_{\nu}^{(0,0)} = m_D^{(0,1)}\frac{1}{M_R}{{m_D^{(0,1)}}^T} + \mathcal{O}\left({\left(m_D^{(0,k)}\right)^2 \over m_{(k)}}\right) 
\end{equation}
 where  higher order corrections are from higher KK states. To fit the neutrino masses and mixing angles we neglect higher order corrections as before.
 Defining $Y'_N = 2 k Y_N$, we have 
 \begin{equation}
  m_{\nu}^{(0,0)} = Y'_N e^{(1-c_L)kR\pi}g_L(\pi R) (M_R^{-1}) Y'_Ne^{(1-c_L)kR\pi}g_L(\pi R)
\label{seesaw} 
\end{equation}
In Table (\ref{bulkmajorana}), we present two sample points one for inverted hierarchy and another for normal hierarchy, which fit the neutrino
masses and mixing angles as well as charged lepton masses  with the accuracy we have specified in section \ref{fitsection}. Both these examples\footnote{These
solutions require that the  profiles of the $N$ fields have very small values on the UV brane.}
have $c_M < c_N$.  The corresponding Yukawa coupling matrices
are presented in Eqs. (\ref{majyuk1},\ref{majyuk2}).
\begin{table}[h]
\caption{Sample points with corresponding fits of observables for Normal and Inverted Hierarchy schemes  in Bulk Majorana case with O(1) Yukawas. The masses are in GeV}
\begin{center}
\begin{tabular}{|c|c|c|}
\hline
Parameter &Normal&Inverted\\
\hline
\hline
$M_{kk}$&161.4&161.4\\
\hline
$c_{M_i}$&0.55&0.55\\
\hline
$g_L^{(1)}(\pi R)$&$3\times 10^{-13}$&$1.2\times 10^{-12}$\\
\hline
$c_{L_1}$ & 0.58&0.59\\ 
\hline
$c_{L_2}$ &0.56& 0.57 \\
\hline
$c_{L_3}$&  0.55&0.55\\
\hline
$c_{E_1}$& 0.735&0.735\\
\hline
$c_{E_2}$ &0.5755&0.575\\
\hline
$c_{E_3}$&0.501& 0.501\\
\hline
$c_{N_i}$&0.58&0.58\\
\hline
$m_e$&$5.09\times10^{-4}$&$5.08\times10^{-4}$\\
\hline
$m_\mu$&0.1055 &0.1055\\
\hline
$m_\tau$&1.77&1.774\\
\hline
$\theta_{12}$&0.58&0.58\\
\hline
$\theta_{23}$&0.80&0.8\\
\hline
$\theta_{13}$&0.13&0.13\\
\hline
$\Delta m_{sol}^2$&$7.8\times10^{-23}$&$7.8\times 10^{-23}$\\
\hline
$ \Delta m_{atm}^2$&$2.4\times 10^{-21}$&$2.4\times 10^{-21}$\\
\hline
\hline
\end{tabular}
\end{center}
\label{bulkmajorana}
\end{table}

Yukawa parameters for inverted hierarchy
 \begin{equation}
 \label{majyuk1}
  Y'_N = \begin{bmatrix}
                          2.73&1.81&.108\\
			  -0.83&0.975&.328\\
			   0.327&-0.679&.182 
                         \end{bmatrix}\;\;Y'_E=\begin{bmatrix}
                                3.44&-0.41&.87\\0.62& 1.583& 0.332\\2.74& 0.55& 2.33
                                               \end{bmatrix}      
 \end{equation}

Yukawa parameters for normal hierarchy
 \begin{equation}
 \label{majyuk2} 
  Y'_N = \begin{bmatrix}
                          2.56&1.69&1.26\\
			  -0.795&0.927&3.89\\
			   0.414&-0.859&2.86 
                         \end{bmatrix}\;\;Y'_E=\begin{bmatrix}
                                2.825&-0.41&.87\\0.62& 1.2008& 0.332\\2.74& 0.55& 2.31
                                               \end{bmatrix}      
 \end{equation}
\subsection{Brane localized Majorana mass term}
Following our discussion with a bulk Majorana mass term, there could be special cases where the Majorana mass term could be localized on either boundary.
In this case the bulk profiles for the right handed singlets $N_i$ remain unchanged. The eigenvalue equations are same as in Eq.(\ref{eigenvalueeqn}).\newline

\subsubsection{UV localized mass term} 

The case with UV localized Majorana mass term was studied in \cite{Huber2,perez}. The action in this case is given as
\begin{equation}
 S_{N}=\int d^4x\int dy~ \sqrt{-g} ~\big(\delta(y)\bar NN^c+ m_D \bar N N + \delta(y-\pi R)Y_N\bar L\tilde H N\big)
\label{majoranaUV}
\end{equation}
where we have expressed $m_M = \delta(y)$.
Substituting the KK expansions from Eq.(\ref{kkexpansion1}), the effective 4-D neutrino mass matrix, in the basis $\chi^T = \{ \nu_L^{(0)}, N_R^{(0)}, N_R^{(1)}, N_L^{(1)}\}$ takes the form

\begin{equation}
\mathcal{L}_m = - {1 \over 2} \chi^T M_N \chi \;\;\;; \;\;\;\; M_N = \begin{bmatrix}
                0&\mathcal{M}^{(0,0)}_\nu&\mathcal{M}^{(0,1)}_\nu&0\\
                \mathcal{M}^{(0,0)}_\nu& M^{Maj}_{\nu^{(0,0)}}&M^{Maj}_{\nu^{(0,1)}}&0\\
		\mathcal{M}^{(0,1)}_\nu&M^{Maj}_{\nu^{(0,1)}}&M^{Maj}_{\nu^{(1,1)}}&M_{KK}\\
                 0&0&M_{KK}&0
               \end{bmatrix}
\label{uvmajorana}
\end{equation}
where $\mathcal{M}^{(0,0)}_\nu$ is defined in Eq.(\ref{mmdirac2}). Let $ f^{(1)}_N(0)$ denote the value of the profile of the first KK mode
of N at the UV brane \textit{i.e,} y=0 and $f_N(0)$, defined in Eq.(\ref{profile}), is the zero mode profile of N evaluated at y=0. The individual elements of Eq.(\ref{uvmajorana}) are then defined as:
 $\mathcal{M}^{(0,1)}_\nu=\frac{v}{\sqrt{2}}\frac{1}{\sqrt{\pi R}}f_N(\pi R)Y_N'$;
$M^{Maj}_{\nu^{(0,0)}} =\frac{1}{\pi R}f_N^2(0)$ ;$M^{Maj}_{\nu^{(0,1)}} = \frac{1}{\pi R}f^{(1)}_N(0)f_N(0)$; $M^{Maj}_{\nu^{(1,1)}} = \frac{1}{\pi R}f^{(1)}_N(0)f^{(1)}_N(0)$ and $M_{KK}$ is the KK mass of
first KK mode of N. The small neutrino masses can be fit by choosing $c_N\sim 0.32$ for which $M^{Maj}_{\nu^{(0,0)}}\sim10^{14} \text{GeV}$. The charged leptons are fit by choosing $c_{L,E} > 0.5.$ This scenario along with flavour implications
has been extensively dealt in \cite{perez}.

\subsubsection{Pure Majorana Case}
An interesting sub case of the Bulk Majorana term would be the situation where $m_D = c_N k =0$. As we have seen from the discussion 
in the previous section, in such a case, the profile equations become oscillatory.   The eigenvalue equations
now take the form: 
\begin{eqnarray}
\partial_y g_L^{(n)}(y)&=&m_ne^\sigma g_R^{(n)}(y)-m_M g_R^{(n)}(y) \nonumber \\
-\partial_yg_R^{(n)}(y)&=&m_ne^\sigma g_R^{(n)}(y)-m_M g_L^{(n)}(y)
\end{eqnarray}
Contrary to the Dirac+ Majorana case of the previous section,  the  above set of equations allow solutions for 
zero modes, $m_{0} = 0$.  The solutions are given as 
\begin{eqnarray}
g_L (y) &=&N \cos(\frac{m_ne^\sigma}{k}-m_My ) \nonumber \\
g_R(y) &=& N\sin(\frac{m_ne^\sigma}{k}-m_My),
\end{eqnarray}
where $N$ is the normalization factor given by $N=\sqrt{\pi Rk}e^{-0.5\sigma(\pi R)}$. These solutions are consistent
with the boundary conditions.  The neutrino mass matrix has a specific structure in this case, as there are
contributions from the first KK mode, which might be important. In the basis, $\chi^T = \{ \nu_L^{(0)}, N^{(0)}, N^{(1)}\}$ 
the mass matrix takes the form 
\begin{equation}
\mathcal{L}_m = - {1 \over 2} \chi^T \mathcal{M} \chi \;\;\;; \;\;\;\; \mathcal{M} =  \left( \begin{array}{ccc} 
0 & m_D^{(0,0)} & m_D^{(0,1)}\\
m_D^{(0,0)} &0& 0 \\
m_D^{(0,1)} & 0 & m_{(1)}
\end{array} \right)
\end{equation}
From the above, we see that at the zeroth level, light neutrino  and singlet neutrinos form a pseudo-Dirac structure, leading to maximal
mixing between these two states.  For the three flavor states, we would have three light states which are sterile. We have not pursued
the phenomenology of this model further.

\section{Lepton Flavor Violation}
\label{seclfv}
We now study lepton flavor violating constraints on the three neutrino mass models considered in the present work. Lepton flavor violation within the
RS framework has been studied in detail in \cite{Agashe}. The localization of the fermions in the bulk at different places leads to non-zero flavour mixing
between the zero mode SM fermions and higher KK states, which contribute to flavor violating processes both at the tree and the loop level.  
The tree level flavor violating decay modes of the form $l_i \to l_j l_k l_k$   are due to non-universal overlap of the zero mode fermions with the Z-boson KK modes.
At the 1-loop level, penguin graphs contribute to rare decays like $l_j \to l_i + \gamma$. The SM states mix with their heavier KK states on the IR brane, and thus may give rise to
significant contributions to dipole processes in particular. The present LFV limits are very strong and are listed in Table[\ref{lfv-tab}]

\begin{table}[htp]
\caption{Present Experimental Bounds on LFV Processes}
\begin{center}
\renewcommand{\arraystretch}{1.5}
\begin{tabular}{ccc}
\hline Process &Experiment & Present upper bound \\ \hline
 BR$(\mu\rightarrow e\, \gamma)$& MEG \cite{meg11, DeGerone:2011fg}& $2.4 \times 10^{-12} $\\
BR$(\mu\rightarrow e\, e\, e)$ & MEG \cite{meg11, DeGerone:2011fg} &  $1.0 \times 10^{-12} $\\
CR$(\mu\rightarrow e\, {\rm in} \, {\bf Ti})$ & SINDRUM-II \cite{Wintz:1996va} & $6.1 \times 10^{-13}$ \\
BR$(\tau\rightarrow \mu\, \gamma)$ & BABAR/Belle \cite{:2009tk} &$4.4 \times 10^{-8}$ \\
BR$(\tau\rightarrow e\, \gamma)$ & BABAR/Belle \cite{:2009tk} & $3.3 \times 10^{-8}$\\
BR$(\tau\rightarrow \mu\, \mu\, \mu)$ &BABAR/Belle \cite{:2009tk} &$2.0 \times 10^{-8}$\\
BR$(\tau\rightarrow e\, e\, e)$ & BABAR/Belle \cite{:2009tk}& $2.6 \times 10^{-8}$ \\ \hline
\label{lfv-tab}
\end{tabular} 
\end{center}
\end{table}
In this section we calculate the Branching fractions for the leptonic FCNC. 
The effective 4-D lagrangian describing $l\rightarrow l'$ process is given by \cite{Agashe}
\begin{eqnarray}
-\mathcal{L_{{\rm eff}}}&=&A_R(q^2)\frac{1}{2m_\mu}\bar{e}_R\sigma^{\mu\nu}F_{\mu\nu}\mu_L+
A_L(q^2)\frac{1}{2m_\mu}\bar{e}_L\sigma^{\mu\nu}F_{\mu\nu}\mu_R \nonumber \\
& &+\frac{4G_F}{\sqrt{2}}\left[a_3(\bar{e}_R\gamma^\mu\mu_R)(\bar{e}_R\gamma_\mu e_R)
+a_4(\bar{e}_L\gamma^\mu\mu_L)(\bar{e}_L\gamma_\mu e_L)\right. \nonumber \\
& &+\left.a_5(\bar{e}_R\gamma^\mu\mu_R)(\bar{e}_L\gamma_\mu e_L)
+a_6(\bar{e}_L\gamma^\mu\mu_L)(\bar{e}_R\gamma_\mu e_R)\right] + {\rm h.c.}
\label{ngeq}
\end{eqnarray}
\subsection{Tree level decays} 
The breaking of the electroweak symmetry at the IR brane mixes the zero mode gauge boson with the higher modes. To parametrize this mixing, let ($Z^{(0)}$, $Z^{(1)}$) and (${Z'}^{(0)}$ ${Z'}^{(1)}$) denote the gauge boson states before and after diagonalisation of the gauge boson
mass matrix respectively. Assuming only one KK mode for simplicity, they are related as \cite{Agashe}
\begin{eqnarray}
 {Z'}^{(0)} = Z^{(0)} + \sqrt{2kR\pi}\frac{m_Z^2}{M_{Z^{(1)}}^2}Z^{(1)} \nonumber \\
{Z'}^{(1)} = Z^{(1)} - \sqrt{2kR\pi}\frac{m_Z^2}{M_{Z^{(1)}}^2} Z^{(0)}\nonumber \\
\end{eqnarray}
where $M_{Z^{(1)}}$ is the mass of first KK excitation of the Z boson. Owing to its flat profile the $Z^{(0)}$ couples universally to all three generations. However, the coupling of ${Z}^{(1)}$, whose profile is peaked near the IR brane, is generation dependent. This coupling depends on the localization of the fermions along the extra-dimension thus giving rise to non-universality. Let $\eta^T =$ 
          \{$e_M$,$\mu_M$, $\tau_M$\}             
  be vector of fermions in the mass basis. Let $a_{ij}^{(1)}$ be a $3\times 3$ matrix which denotes the coupling of
SM fermions in the mass basis to ${Z'}^{(1)}$. It is given as
\begin{equation}
 a^{(1)ij}_{L,R}=g_{L,R}~ \bar\eta_{L,R}.D_{L,R}^\dagger.\begin{bmatrix}
                I_e&0&0\\0&I_\mu&0\\0&0&I_\tau 
                \end{bmatrix}.D_{L,R}.\eta_{L,R}~\slashed{{Z'}}^{(1)}
\end{equation}
where $g_{L,R}$ is the SM coupling, $D_{L,R}$ are $3\times 3$ unitary matrices for rotating the zero mode (SM) fermions from the flavour basis to the mass basis. \textit{I} is the overlap of the profiles of two zero mode fermions and first KK gauge boson. It is given by
\begin{equation}
 I(c)=\frac{1}{\pi R}\int_0^{\pi R}dy e^{\sigma(y)}(f_i^{(0)}(y,c))^2\xi^{(1)}(y)
\end{equation}
$\xi^{(1)}(y)$ denotes the profile of the first KK gauge boson. It is plotted as a function of a generic bulk mass parameter c in Fig.[\ref{overlap}]. As we can see from this figure, the overlap
function $I(c)$ becomes universal for $c>0.5$ and for $c\lesssim-15$. The off diagonal elements of $a_{ij}^{(1)}$ represent the flavour
violating couplings.  
The contribution to $l_i \to l_j l_k l_k $  from direct $Z^{(1)}$ exchange is suppressed compared to that of ${Z}^{(0)}$. The contributions to the coefficients $a^{ij}_{3,.,6}$ in Eq.(\ref{ngeq}) 
due to the  flavour violating coupling of ${Z}^{(0)}$ as well as direct ${Z}^{(1)}$ exchange are given as
\begin{eqnarray}
 a^{ij}_3 &=& -2g_R(\sqrt{2kR\pi}-I_j)\frac{m_Z^2}{M_{Z^{(1)}}^2} a^{(1)ij}_{R}\\ \nonumber
a^{ij}_4 &=& -2g_L(\sqrt{2kR\pi}-I_j)\frac{m_Z^2}{M_{Z^{(1)}}^2} a^{(1)ij}_{L}\\ \nonumber
a^{ij}_5 &=& -2g_L(\sqrt{2kR\pi}-I_j)\frac{m_Z^2}{M_{Z^{(1)}}^2} a^{(1)ij}_{R}\\ \nonumber
a^{ij}_6 &=& -2g_R(\sqrt{2kR\pi}-I_j)\frac{m_Z^2}{M_{Z^{(1)}}^2} a^{(1)ij}_{L}
\end{eqnarray}
The Branching fractions for the tree level decays are given as \cite{Agashe}
\begin{eqnarray}
BR(\mu \rightarrow 3e) &=& 2\left(|a^{\mu e}_3|^2+|a^{\mu e}_4|^2\right)+|a^{\mu e}_5|^2+|a^{\mu e}_6|^2 \nonumber \\
BR(\tau \rightarrow 3\mu) &=& \left\{2\left(|a^{\tau \mu}_3|^2+|a^{\tau \mu}_4|^2\right)+|a^{\tau \mu}_5|^2+|a^{\tau \mu}_6|^2 
   \right\} BR(\tau \to e\nu\nu) \nonumber \\
BR(\tau \rightarrow 3e) &=& \left\{2\left(|a^{\tau e}_3|^2+|a^{\tau e}_4|^2\right)+|a^{\tau e}_5|^2+|a^{\tau e}_6|^2 
   \right\} BR(\tau \to e\nu\nu)\nonumber \\
BR(\tau \rightarrow \mu ee) &=& \left\{|a^{\tau \mu}_3|^2+|a^{\tau \mu}_4|^2+|a^{\tau \mu}_5|^2+|a^{\tau \mu}_6|^2 
   \right\} BR(\tau \to e\nu\nu)\nonumber \\
BR(\tau \rightarrow e\mu\mu) &=& \left\{|a^{\tau e}_3|^2+|a^{\tau e}_4|^2+|a^{\tau e}_5|^2+|a^{\tau e}_6|^2
   \right\} BR(\tau \to e\nu\nu).
\label{bfracs}
\end{eqnarray}

\begin{figure}
 \includegraphics[width=0.50\textwidth,angle=0]{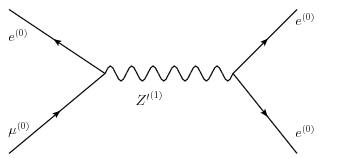}
\caption{Tree level contribution to $\mu\rightarrow eee$ due to exchange of ${Z'}^{(1)}$. The effective ${Z}^{(0)}$ contribution is proportional to this graph.}
\label{trilepton1}
\end{figure}

\noindent
Similarly, the relevant quantities for $\mu \to e$ conversion in Ti are given as: 
 \begin{eqnarray}
a_{L,R}^{\mu e} &=& -\sqrt{2kR\pi}\frac{m_Z^2}{M_{Z^{(1)}}^2} a^{(1)\mu e}_{L,R}\nonumber \\
 \hspace{0.8cm}BR(\mu\rightarrow e)~\text{ in Nuclei} &=& \frac{2p_eF_p^2E_eG_F^2m_\mu^3\alpha^3Z^4_{eff}Q_N^2}{\pi^2Z\Gamma_{capt}}[|a_R^{\mu e}|^2+|a_L^{\mu e}|^2] 
\end{eqnarray}
where $p_e\sim E_e \sim m_\mu$. $G_F$ is the Fermi constant, $\alpha$ is the electromagnetic coupling. The most stringent constraint for
$\mu-e$ conversion comes from Titanium ($Ti^{48}_{22}$). Atomic constants are defined as: $Q_N = v^u(2Z+N) + v^d(2N+Z)$ with N being the neutron number, $Z_{eff}=17.61$, form factor $F_p=0.55$, $\Gamma_{capt}=2.6\times 10^6$ $s^{-1}$ for
Titanium \cite{Chang}.

\subsection{Dipole Transition $l_j\rightarrow l_i\gamma$}
\begin{figure}
 \includegraphics[width=0.50\textwidth,angle=0]{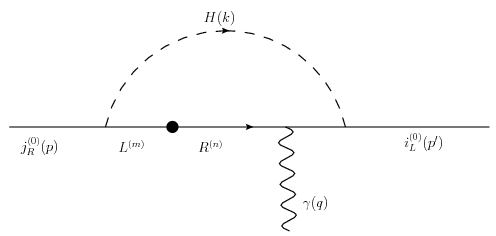}
\caption{Higgs mediated $j\rightarrow i\gamma$. The dot represents the mass insertion. Flavour indices have been suppressed in the internal charged KK lines. (L,R) represents
the KK modes corresponding to the left and right chiral zero modes.}
\label{higgsdipole1}
\end{figure}

The dominant graph is due to scalar exchange in the loop. One of them is due to Higgs exchange as shown in Fig.[\ref{higgsdipole1}]. The amplitude for this process is given as
\begin{equation}
 M_{j\rightarrow i\gamma} = \sum _{n,m}\int \frac{d^4k}{(2\pi)^4} \bar u_i(p'){\Lambda^i_{L}}^\dagger{Y'_E}\frac{\hat p ' +M_n}{\hat p'^2- M_n^2}e\gamma^\mu\frac{\hat p  +M_n}{\hat p^2- M_n^2}v{Y'_E}^\dagger\frac{\hat p  +M_m}{\hat p^2- M_m^2}Y'_E \Lambda^j_{R}u_i(p)\frac{1}{k^2-m_H^2}\epsilon_\mu 
\label{amplitude}
\end{equation}
where $\hat p = p - k$, $\hat p' = p' - k$ and $q=p-p'$. $\Lambda^i_{L,R}=F_{L,R}^iD_{L,R}$ and $M_{n}$ denotes the mass of the $n^{th}$ mode KK fermion. $F_{L,E}$ is a function of bulk masses which are taken to be diagonal in the flavour space. It is given as 
\begin{equation}
 F_{L,R}=\begin{bmatrix}
          f_{c_{L_1},c_{E_1}}(\pi R)&0&0\\0&f_{c_{L_2},c_{E_2}}(\pi R)&0\\0&0&f_{c_{L_3},c_{E_3}}(\pi R)
         \end{bmatrix}
\end{equation}

 The amplitude for Eq.(\ref{amplitude}) can be rewritten as
\begin{equation}
 M(j\rightarrow i\gamma) = (eD_L^\dagger F_LY'_E{Y'_E}^\dagger v  Y'_E F_R D_R)_{ij}J(\hat p, \hat p',q)
\label{dipole}
\end{equation}
The expression $J(\hat p, \hat p',q)$ is the momentum integral in Eq.(\ref{amplitude}). It is log divergent owing to a double-independent sum over two KK modes. We regularise it using a cutoff of $\Lambda \sim 4\pi M^{(1)}_{kk} \sim 15$ TeV.
The other dominant contribution is due to Fig.[\ref{higgsdipole2}] is discussed in Appendix[\ref{dipolegraphs}].
The Branching fraction for the dipole decays $l_j\rightarrow l_i\gamma$ is given as
\begin{equation}
 BR(l_j\rightarrow l_i\gamma)=\frac{12\pi^2}{(G_Fm_j^2)^2}(A_L^2+A_R^2)
\end{equation}
where the coefficient due to Figs.[\ref{higgsdipole1},\ref{higgsdipole2}] is given as
\begin{equation}
 A_L=2\frac{em_j}{16\pi^2}\frac{1}{M_{KK}^2}\frac{v}{\sqrt{2}}D_L^\dagger F_L(Y'_N{Y'_N}^\dagger+Y'_E{Y'_E}^\dagger)Y_E F_R D_R
\end{equation}
and $A_R=A_L^\dagger$. The other dipole contributions 
are discussed in Appendix[\ref{dipolegraphs}].
   We now proceed to discuss the LFV rates for the mass models discussed in Section[\ref{fitsection}]. The quantities, like the KK masses of fermions, the rotation matrices $D_{L,R}$ etc. which determine the LFV rates
are functions of the bulk mass parameters. We compute these quantities for each point of the best fit parameter space obtained earlier for the LHLH and the Dirac case and use it to constrain
the parameter space from flavour violation.

\subsection{LHLH Case} 
\begin{figure}
 \includegraphics[width=0.50\textwidth,angle=0]{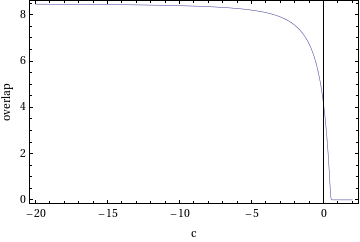} 
\caption{Coupling of two zero mode fermions to $Z_1$ as a function of bulk mass parameter \cite{Gher}.}
\label{overlap}
\end{figure}

\begin{figure}[htp]
\begin{center}
\includegraphics[angle=0,width=.5\textwidth]{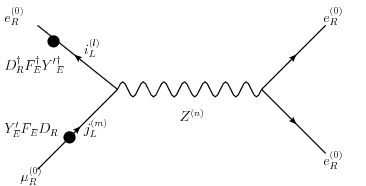} 
\end{center}
\caption{Additional tree level contribution to $\mu\rightarrow eee$. For a fairly degenerate bulk doublet mass in LHLH case this contribution is negligible. For the Dirac
case this graph receives wave function suppression in addition to the KK scale suppression.}
\label{trilepton2}
\end{figure}

The contributions to trilepton decays from graphs like Fig.[\ref{trilepton1}] are highly suppressed in the parameter space of interest. 
This is because the couplings of the zero mode fermions to the KK gauge boson become universal for the fermions sufficiently localized
towards IR and UV branes, as can be seen in Fig.[\ref{overlap}].   However, there could be other potentially large contributions. This
comes from the large mixing between zero mode charged singlet states and the first KK modes of the  lepton doublets; the corresponding
Yukawa  coupling is very large due to the  large negative $c_{E}$ values.  Example of such a graph is  shown in Fig.[\ref{trilepton2}]. 
Exact value of the contribution, of course depends on the values of $D_{L,R}$ and other parameters. We have not considered these
graphs in the present work. We note that  for a fairly degenerate bulk doublet masses, ($c_{L_i}$), the  combination of the matrices
which enter in these graphs  are aligned with the zero mode mass matrix for charged leptons. The best parameter space does
contain such regions where all the $c_{L_i}$ are degenerate. We found several examples of that kind. 
Another potential problem with the highly localized IR charged singlets, is the shift in the universal coupling constant $g_R$.
This could effect  $Z\rightarrow ll$ branching fractions. Models with custodial symmetries or very heavy  KK gauge bosons  
could avoid this problem. We have not addressed this issue here. 
  
Finally, contribution to $l_j\rightarrow l_i\gamma$ due to loop diagrams of the form in Fig.[\ref{dipole3}] are heavily suppressed owing to the heavy
 KK  mass scales corresponding to the charged singlets. The corresponding masses are in shown in Table[\ref{lhlhnormalrange}]. 
 Additionally, the large effective 4-D Yukawa couplings of the charged singlets to the KK modes make it difficult to apply techniques of perturbation theory to calculate graphs like those in Fig.[\ref{higgsdipole1},\ref{higgsdipole2}].

\subsection{Constraints on Dirac Neutrinos}
\begin{figure}[htp]
\begin{tabular}{cc}
\includegraphics[width=0.50\textwidth,angle=0]{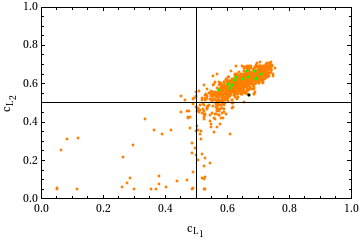} &
\includegraphics[width=0.50\textwidth,angle=0]{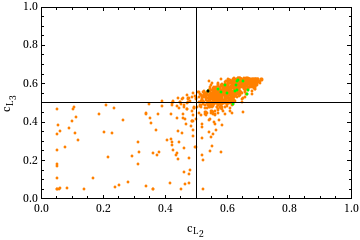} \\
\includegraphics[width=0.50\textwidth,angle=0]{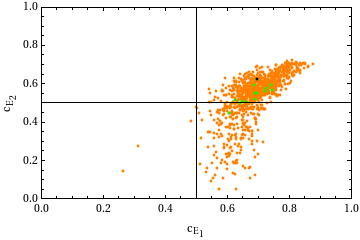} &
\includegraphics[width=0.50\textwidth,angle=0]{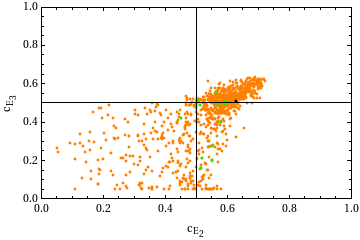}
\end{tabular}
 \caption{The black dot and the green region represent the parameter space permitted by tree-level constraints for a KK gauge boson scale of 1920 and 3000
GeV respectively}
\label{treelevel}
\end{figure}

The Dirac case gives a good fit to the leptonic data for a reasonable choice of $\mathcal{O}$(1) parameters. However, the parameter space is strongly
constrained from flavour considerations. In the parameter space of interest the dominant contribution to tree-level decays comes from Fig.[\ref{trilepton1}]. The parameter space of the bulk doublets and charged singlets consistent with tree level
contribution is shown in Fig.[\ref{treelevel}]. The lightest  $M_{Z^{(1)}}$  mass required to satisfy all constraints from tree-level processes $\sim 1.9~\text{TeV}$. Fig.[\ref{treelevel}] shows the points within the  best fit parameter space
consistent with all constraints from tree-level processes.  As can be seen from the figure, very few points pass the constraints. 
The black point is allowed for a  KK gauge boson scale of $1.9$ TeV, where as the
green points are for mass of 3 TeV.
The constraints from dipole processes are far more severe. Corresponding to the $c_{L,E}$ values in the best fit parameter space,  the mass of the first
KK excitation of the leptons varied from approximately 850 GeV to 1400 GeV as presented in Table (\ref{diracnormalrange}). 
We found no points which satisfied the constraints from $\mu\rightarrow e\gamma$, $\tau \rightarrow e\gamma$ and $\tau \rightarrow\mu\gamma$ 
simultaneously. The constraint from $\mu\rightarrow e\gamma$ was most severe and required a KK fermion mass scale $\mathcal{O}$(10) TeV to suppress it
below the experimental limit given in Table [\ref{lfv-tab}].  

\subsection{Constraints on scenarios with bulk Majorana mass}
The tree-level decays only constrain the parameter space of the bulk doublets and charged singlets as seen in Fig.[\ref{treelevel}]. Since, the
charged lepton mass fitting is independent of any right handed neutrino parameter, the constraints coming from tree-level decays in the Dirac
case are applicable in this case as well.

 The contribution to dipole decays of the form $l_j\rightarrow l_i\gamma$ due to charged Higgs shown in Fig.[\ref{higgsdipole2}] is small. This is because, as shown in Table[\ref{bulkmajorana}], $g_L^{(1)}(\pi R)$ is required to be small to fit neutrino masses.
Thus, the dominant contribution to dipole decays in this case is due to Higgs exchange diagram shown in Fig.[\ref{higgsdipole1}].  
They are calculated for the both the normal and inverted hierarchy cases presented earlier and are given in Table[\ref{bulkmajdipole}]. The branching fractions are evaluated for $M_{KK}\sim 1250$ GeV which is
the first KK scale of the doublet.
\begin{table}[htdp]
\caption{BR for dipole decays for the case with bulk Majorana mass}
\begin{center} 
\begin{tabular}{|c|c|c|c|}
\hline
Hierarchy&BR($\mu \rightarrow e\gamma$)&BR($\tau \rightarrow \mu\gamma$)&BR($\tau \rightarrow e\gamma$) \\
\hline
\hline
Inverted & $2.4\times 10^{-5}$ &$1.9\times 10^{-5}$ & $7.6\times 10^{-6}$\\
Normal & $1.4\times 10^{-5}$&$3.4\times 10^{-5}$&$1.3\times 10^{-5}$\\
\hline 
\end{tabular}
\end{center}
\label{bulkmajdipole}
\end{table}

\section{Minimal Flavor Violation(MFV)} 
\label{secmfv}

From the discussion above it is clear that lepton flavor violating
constraints are strong on  RS models with fermions
localized in
 bulk and Higgs localized on the IR brane.   In the Dirac and the Bulk
Seesaw case  flavor violation rules out most of the
`best fit' parameter space.
One option to evade these bounds would be to increase  the scale of KK
masses. As we have seen in the LHLH case,  the fits indicate to the
highly hierarchal spectrum with KK masses of the $\mathcal{O}(10^{2})$ TeV
for the singlet charged leptons, the  flavor violating  amplitudes
are highly suppressed and thus do  not put severe constraint  on the model. However, the
Dirac and the Majorana cases whose best fit regions have lighter
KK spectrum would essentially be ruled out.  The misalignment between
the Yukawa coupling matrix and bulk mass terms which determine
the profile is the cause of the large flavor violating transitions
leading to strong restrictions on these models. 
In \cite{a4delaunay} the authors imposed discrete symmetries to constrain Flavour Changing Neutral Currents (FCNC).  
In this work we adopt the Minimal Flavour violation  ansatz which reduces the
misalignment by demanding an alignment between the Yukawa matrices and the
bulk parameters. 

The ansatz of Minimal Flavour violation was first proposed for the hadronic sector \cite{mfv1}. It proposes that new physics adds no new flavor structures and thus
 entire flavor structure in Nature is  determined by the Standard Model Yukawa couplings. In the leptonic sector, MFV  in not uniquely defined due to the
 possibility of the seesaw mechanism. Several schemes of leptonic minimal flavor violation are possible \cite{cirigliano}.

The proposal to use the MFV hypothesis in RS was
first introduced in \cite{Fitzpatrick} in the quark sector. There were
subsequent extensions in the leptonic sector by  \cite{perez,Chen}.
 The MFV ansatz assumes that the  Yukawa couplings are the only
sources of flavor violation. In the RS setting this would require that
the  bulk mass terms should now be expressed  in terms of the Yukawa
couplings \cite{Fitzpatrick}.  The exact expression would depend on the
 particle content and the flavor symmetry assumed.

\subsection{Dirac Neutrino Case}
In the presence of right handed neutrinos the flavour group is $SU(3)_L\times SU(3)_{E}\times SU(3)_{N}$; the lepton number is conserved. 
 The $Y_E$ transforms  as $Y_E \rightarrow (3,\bar 3,1)$ 
and $Y_N$ transforms as $Y_N\rightarrow (3,1,\bar 3)$.
 The Yukawa couplings  are aligned with the five dimensional  bulk mass matrices. The bulk masses can be expressed in terms of the Yukawas as
\begin{equation}
 c_L=a_1I+a_2 {Y'}_EY'^\dagger_E+a_3Y'_NY'^\dagger_N \;\;\;\;\;c_E=bY'^\dagger_E Y'_E \;\;\;\;\;c_N=cY'^\dagger_N Y'_N
\end{equation}
where a,b,c $\in \Re$ and $Y'_{E,N}$ are as defined earlier as $Y'_{E,N}=2kY_{E,N}$.
Owing to the flavor symmetry we work in a basis in which $Y'_E$ is diagonal. We then rotate $Y'_N$ by the PMNS matrix \textit{i.e,}
writing $Y'_N\rightarrow V_{PMNS} \text{Diag}(Y'_N)$ where the $\text{Diag}(Y'_N)=\text{Diag}(0.709,0.709,0.75)$. The $c_L$ value chosen is $0.5802$ for all three generations. The $c_{N}$ values chosen are respectively 1.17241, 1.172, 1.311 respectively.
The bulk singlet mass parameters are $c_E =(0.7477,0.58059,0.401)$

The simplest Yukawa combination transforming as (8,1,1) under the flavour group is given as
\begin{equation}
 \Delta = Y'_N Y'^\dagger_N
\end{equation}
Thus the BR for $\mu\rightarrow e\gamma$, which is the most constrained is given as \cite{perez}
\begin{equation}
 BR(\mu\rightarrow e\gamma) = 4\times 10^{-8}~(Y'_N Y'^\dagger_N)^2_{12}~\Big(\frac{3\text{TeV}}{M_{KK}}\Big)^4
\end{equation}

\begin{equation}
 Y'_{N}= \begin{bmatrix}
          0.586033& 0.383951& 0.115044\\ -0.335962& 0.370429& 0.53165\\
0.215349&-0.466953 & 0.516346
          \end{bmatrix}
\end{equation}
The (1,2) element of $\Delta$ which is responsible for $\mu\rightarrow e\gamma$ is 0.006 which gives a contribution of $1.44\times 10^{-12}$, for a fermion
 KK mass of around 3 TeV.

\subsection{Bulk Majorana mass term}
Owing to the presence of a bulk Majorana mass term,  we choose the flavour group  for the lagrangian in Eq.(\ref{majorana}) is $SU(3)_L\times SU(3)_E\times O(3)_N$. $Y_E$ transforms as $Y_E \rightarrow (3, \bar 3,1)$ 
 and $Y_N$ transforms as $Y_N\rightarrow (3,1,3)$. The bulk Majorana term $\bar N^cN$ transforms as $(1,1,6)$ under this flavour group. In terms of the dimensionless Yukawa couplings, $Y'_{E,N}$ the  bulk mass parameters can be expressed as
\begin{equation}
 c_L=a_1I+a_2 {Y'}_EY'^\dagger_E+a_3Y'_NY'^T_N \;\;\;\;\;c_E=1+bY'^\dagger_E Y'_E \;\;\;\;\;c_N=1+cY'^T_N Y'_N\;\;\;\;\;c_M=dI_{3\times3}
\end{equation}

where a,b,c,d $\in \Re$. $c_M=0.55$ and $c_N=0.58$ are chosen for the right handed neutrino bulk mass parameters. The value of profiles for the singlets are chosen appropriately at the boundary so as to fit the neutrino data
using the $\mathcal{O}$(1) Yukawa couplings. As before we work in a basis in which $Y'_E$ is diagonal. In this basis $Y'_N = V_{PMNS}\text{Diag}(Y'_N)$. This removes the dominant contribution to dipole decays
due to the Higgs exchange in Fig.[\ref{higgsdipole1}]. The contribution due to Fig.[\ref{higgsdipole2}] is very small owing to wavefunction suppression of the 
singlet neutrinos. Thus, we see that the MFV ansatz is successful in suppressing FCNC's for both the Dirac and the bulk Majorana case.


\section{Summary and Outlook}
\label{secfinal}
Understanding neutrino masses and mixing is an important  aspect of most physics beyond the Standard Model frameworks. The 
Randall-Sundrum setup while solving the hierarchy problem could also form a natural setting to explain flavour structure of the
Standard Model Yukawa couplings. The quark sector has already been explored in this context in detail. While there have been 
several analysis in the leptonic sector, in the present work we have tried to explore the  same in a comprehensive manner, filling 
the gaps wherever we found it necessary. Our aim had been to determine quantitavely the parameter space of both the
$\mathcal{O}(1)$ (dimensionless)  Yukawa couplings as well as the bulk mass parameters which can give good fits to the leptonic data. 

We have concentrated on the RS setup with  the Higgs field localized on the IR boundary.  We have considered 
 three cases of neutrino mass models (a) The LH LH higher dimensional operator (b) The Dirac case and the
(c) Majorana case.  The LHLH fits require large negative c-parameters which reflect the composite nature of the 
charged singlets. There is some parameter space in this case where the flavor constraints are weak. 
However, the model has very large  effective 4-D Yukawa couplings between the zero mode SM fermions and
the KK fermions, which makes it unattractive from perturbation theory point of view. 
 We have also presented the distributions of the Yukawa couplings in the best fit region.  Most of the individual Yukawa 
 couplings are concentrated on the higher side of the $\mathcal{O}(1)$ range we have chosen.  
 The Dirac and Majorana cases offer large parameter space without the need
of large hierarchies in the $c$ parameters. We have also presented the distribution of the Yukawa couplings in 
the Dirac case. We could not find strong
 correlations between the Yukawa couplings and the $c$-parameters. 
There are  strong constraints from the lepton  flavor violating rare
processes. These can be circumvented by a suitable choice of Yukawa couplings and c-parameters
guided by the MFV ansatz. The Majorana case, in particular allows for several classes of MFV schemes, which 
will be explored in an upcoming publication \cite{ourrs2}.

While we restricted ourselves to the Higgs located on the IR brane, it can also be
allowed to propagate in the bulk.
Lepton flavor violating amplitudes however are now cut-off independent, which makes the computations more
predictive. But with the Higgs boson
in the bulk one has to invoke other scenarios like supersymmetry  to solve the hierarchy problem. 
 \newline

 \noindent
\textbf{Acknowledgments}\\  
\newline We thank Bhavik Kodrani for important and interesting inputs. We appreciate D. Chowdhury and R. Garani's help with the numerics. We also thank
V.S. Mummidi for carefully reading the manuscript. SKV acknowledges support from DST Ramanujam fellowship  SR/S2/RJN-25/2008 of Government of India.

 \appendix

\section{Inverted Mass fits}
\label{invertedfits}

 We present the results of the scan performed for inverted hierarchy for both the LLHH and the Dirac case.
 In the case for the normal hierarchy it was easier to find c values and order one
 Yukawa entries which satisfied all constraints. However, the choice of these parameters which fits the data in the 
 inverted case is very subtle. This is because one requires two large mass eigenvalues in the inverted
 case which must satisfy the $\Delta m^2_{sol}$ constraint. This requires a very careful choice of order one Yukawa parameters.
 The parameter space for c values does not differ much between the normal and the inverted case. For the case of inverted hierarchy, we choose
points which satisfy $0<\chi^2<10$. For the Dirac case we performed a scan only for $c>0.5$.\newline
 
(A) \textbf{LHLH case} \newline                                                                                                             
\begin{table}[htdp]
 \caption{Sample points for Inverted Hierarchy  in LHLH case with O(1) Yukawas. The masses are in GeV}
 \begin{center}
 \begin{tabular}{|c|c|c|}
 \hline
 Point  & A&B\\  
 \hline
 \hline
 $\chi^2$&7.48&6.61 \\
 \hline
 $c_{L_1}$& 0.8967&0.9162\\
 \hline
 $c_{L_2}$&0.8983 &0.8920\\
 \hline
 $c_{L_3}$&0.8913& 0.8945 \\
 \hline
 $c_{E_1}$&-3758.1502& -2099.8993 \\
 \hline
 $c_{E_2}$&-6005847.4955&-552577.8188 \\
 \hline
 $c_{E_3}$&-32730342.0982&-23953472.2265\\
 \hline
 $m_e $&$5.11\times10^{-4}$&$5.09\times10^{-4}$ \\
 \hline 
 
 $m_\mu$& 0.1056&0.1056\\
 \hline
 $m_\tau$&1.775&1.755\\                                    
  \hline
     $\theta_{12}$&0.584&0.55\\
 \hline
  $\theta_{23}$&0.829&0.875\\
 \hline
    $\theta_{13}$&0.148&0.160\\
 \hline
  $\delta m_{sol}^2$&$7.49\times 10^{-23}$&$7.46\times 10^{-23}$\\
 \hline
   $ \delta m_{atm}^2$&$  1.90\times 10^{-21}$&$  2.7\times 10^{-21}$\\                            
 \hline 
 \hline 
 \end{tabular}
 \end{center}
 \label{lhlhobs}
 \end{table}
 
 Yukawa for Point A
 \begin{equation}
  Y'_E=\begin{bmatrix}
              0.8249    &          0.8516      &        1.1111      \\        1.3600     &         1.5956      &        1.8402       \\       3.5831      &        3.5664     &         2.9092
  \end{bmatrix} \;\;\;; \;\;\; \kappa'=\begin{bmatrix}
             -3.5528     &         2.6612       &       1.4503      \\        2.6612       &       3.8149      &        1.2903      \\        1.4503     &         1.2903       &      -0.6682
 \end{bmatrix}
 \end{equation}
 
 Yukawa for Point B
 \begin{equation}
  Y'_E=\begin{bmatrix}
              2.5874        &      0.5123      &        3.6064      \\        3.9696     &         2.4876     &         1.9903      \\        3.8604     &         1.1438     &         3.9712
  \end{bmatrix} \;\;\;; \;\;\; \kappa'=\begin{bmatrix}
             -3.6860      &       -3.6778      &        3.9987       \\      -3.6778     &         2.1362    &          3.3252        \\      3.9987       &       3.3252       &      -0.8497
 \end{bmatrix}
 \end{equation}

 (B) \textbf{Dirac Case}\newline
 \begin{table}[h]
 \caption{Sample points for Inverted Hierarchy  in Dirac case with O(1) Yukawas. The masses are in GeV}
 \begin{center}
 \begin{tabular}{|c|c|c|}
 \hline
 Parameter &Point A&Point B\\
 \hline
 \hline
 $\chi^2$&0.30&8.04\\
 \hline
 $c_{L_1}$ & 0.5565&0.51\\ 
 \hline
  $c_{L_2}$ &0.5556& 0.5316 \\
 \hline
  $c_{L_3}$&  0.5433&0.5012\\
 \hline
 $c_{E_1}$& 0.7681&0.8092\\
 \hline
 $c_{E_2}$ &0.6186&0.6498\\
 \hline
 $c_{E_3}$&0.5044& 0.5674\\
 \hline
 $c_{N_1}$&1.2450&1.2765\\
 \hline
 $c_{N_2}$&1.2421&1.2755\\
 \hline
 $c_{N_3}$&1.2546&1.2941\\
 \hline
 $m_e$&$5.1\times10^{-4}$&$5.08\times10^{-4}$\\
 \hline
 $m_\mu$&0.1055&0.1055\\
 \hline
 $m_\tau$&1.769&1.81\\
 \hline
 $\theta_{12}$&0.59&0.59\\
 \hline
 $\theta_{23}$&0.80&0.72\\
 \hline
 $\theta_{13}$&0.155&0.152\\
 \hline
 $\delta m_{sol}^2$&$7.49\times10^{-23}$&$7.48\times 10^{-23}$\\
 \hline
 $ \delta m_{atm}^2$&$2.40\times 10^{-21}$&$2.16\times 10^{-21}$\\
 
 \hline
 \hline
  \end{tabular}
 \end{center}
 \label{diracinverted}
 \end{table}
 
 Yukawa for Point A
 \begin{equation}
  Y'_E=\begin{bmatrix}
               2.2645        &      2.7691      &        0.4272    \\          1.0499      &       -3.6695     &        -1.0818       \\      -2.2402       &      -0.5400      &       -1.9176
  \end{bmatrix} \;\;\;; \;\;\; Y_N'=\begin{bmatrix}
              -0.2202      &       -2.3054        &      1.5602       \\       3.4794     &        -2.2140     &         0.2302    \\         -2.0676        &     -1.7529      &        0.7888
 \end{bmatrix}
 \end{equation}
 
 Yukawa for Point B
 \begin{equation}
  Y'_E=\begin{bmatrix}
              -3.7916     &        -0.3960      &       -2.5573      \\        1.2699     &        -2.3757       &       3.2167       \\      -3.5010        &      3.4430       &       2.8224
  \end{bmatrix} \;\;\;; \;\;\; Y_N'=\begin{bmatrix}
              -3.9443      &       -0.9714      &        0.1848       \\      -2.5788     &         0.2609         &     3.3684      \\        0.5020          &   -3.0268      &       -3.1765
 \end{bmatrix}
 \end{equation}
 
 \begin{figure}[H]
 \begin{tabular}{cc}
  \includegraphics[width=0.50\textwidth,angle=0]{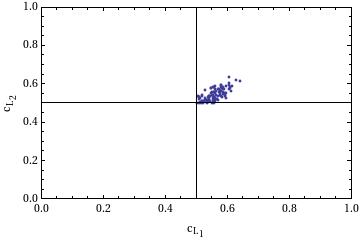}&
  \includegraphics[width=0.50\textwidth,angle=0]{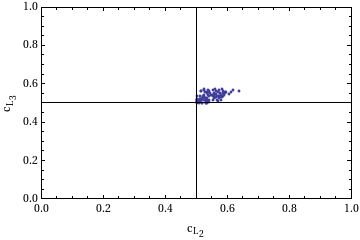}
 \end{tabular}
  \caption{The plot
 represents the parameter space for the bulk masses of charged doublets for inverted hierarchy
 }
 \end{figure}
 
 \begin{figure}[H]
 \begin{tabular}{cc}
  \includegraphics[width=0.50\textwidth,angle=0]{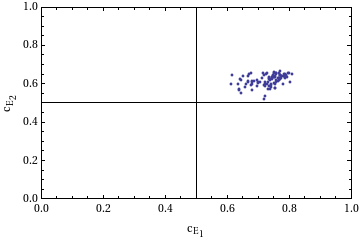}&
  \includegraphics[width=0.50\textwidth,angle=0]{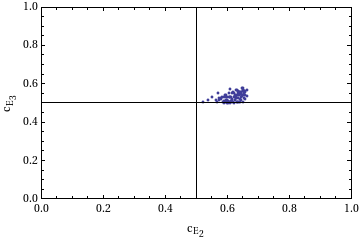}
 \end{tabular}
  \caption{The plot
 represents the parameter space for the bulk masses of charged singlets for inverted hierarchy
 }
 \label{diracinv}
 \end{figure}
 
 \begin{figure}[H]
 \begin{tabular}{cc}
  \includegraphics[width=0.50\textwidth,angle=0]{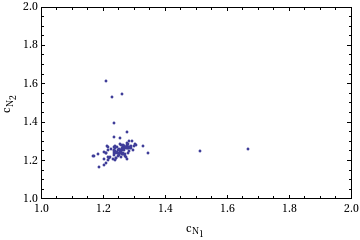}&
  \includegraphics[width=0.50\textwidth,angle=0]{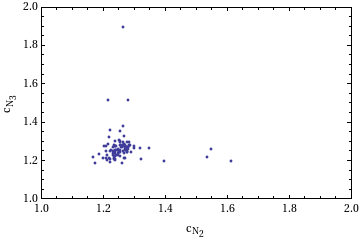}
 \end{tabular}
  \caption{The plot
 represents the parameter space for the bulk masses of neutrino singlets for inverted hierarchy
 }
 \end{figure}

\section{Amplitudes for dipole transitions}
\label{dipolegraphs}
\begin{figure}[htp]
\begin{center}
\includegraphics[angle=0,width=.6\textwidth]{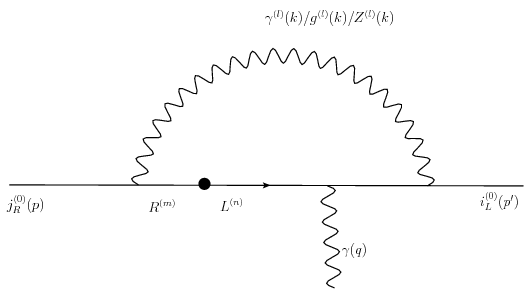} 
\end{center}
\caption{Contribution to the dipole graph due exchange of KK gauge bosons and charged KK fermion lines.}
\label{dipole3}
\end{figure}

In this section we review the other potential contributions to the dipole processes $i\rightarrow j\gamma$ \newline

a) \textbf{Internal flip in neutrino KK line in Dirac Case} \newline
\newline This contribution is absent for the LHLH case as it involves neutral internal KK lines corresponding to the right handed neutrino. In the unitary gauge the charged Higgs
is nothing but the longitudnal component of the W boson.
\begin{figure}[htp]
\begin{center}
\includegraphics[angle=0,width=.6\textwidth]{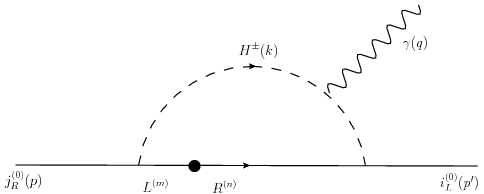} 
\end{center}
\caption{``Charged'' Higgs mediated $j\rightarrow i\gamma$. The dot represents the mass insertion. Flavour indices have been suppressed in the internal neutral KK lines. (L,R) represents
the KK modes corresponding to the left and right chiral zero modes respectively.}
\label{higgsdipole2}
\end{figure}
This displays a similar divergence to Fig.[\ref{higgsdipole1}] owing to the presence of double KK sum.

 \begin{equation}
 M_{j\rightarrow i\gamma} =(F_LY'_N{Y'_N}^\dagger e\frac{v}{\sqrt{2}}  Y'_E F_E)_{ij}\int\sum_{n,m}\frac{d^4k}{(2\pi)^4}\bar u_i(p')(2k^\mu-q^\mu)\frac{(\hat p' +M_n)}{\hat p'^2-M_N^2)}\frac{\hat p +M_n}{\hat p^2-M_m^2}\frac{1}{k^2-m_H^2}\frac{1}{(k-q)^2-m_H^2}u_j(p) 
\end{equation}

b) \textbf{Gauge contribution}\newline
\newline Additional contributions arise due to KK gauge bosons in the loop as shown in Fig.[\ref{dipole3}]

The amplitude for Fig.[\ref{dipole3}] is given as 
\begin{equation}
 M_{j\rightarrow i\gamma} = (A^{0,n,l}\frac{v}{\sqrt{2}} {Y'_E} A^{0,m,l})_{ij}\sum _{n,m}\int \frac{d^4k}{(2\pi)^4} \bar u_i(p')\frac{\hat p ' +M_n}{\hat p'^2- M_n^2}e\gamma^\mu\frac{\hat p  +M_n}{\hat p^2- M_n^2}\frac{\hat p ' +M_m}{\hat p'^2- M_m^2}u_j(p)\frac{1}{k^2-m_H^2} 
\label{dipolegauge}
\end{equation}

where $A^{0,n,l}$ represents the coupling of zero mode fermion to $n^{th}$ mode fermion and $l^{th}$ mode gauge boson. The contribution from
this sector is suppressed in both the Dirac and LHLH case in the parameter space under consideration. \newline

\bibliographystyle{ieeetr}

       \bibliography{RS.bib}

\end{document}